%% file: book.tex
\documentclass[runningheads,fleqn]{svmult}

\usepackage{makeidx}   
\usepackage{graphicx}  
\usepackage{subeqnar}  
\usepackage{multicol}  
\usepackage{physmult}  
\makeindex             

\begin{document}
\title*{Ferromagnetism in (III,Mn)V Semiconductors}
\toctitle{yes, yes, this is the title, but oops, \protect\newline this is a new
line}
%
\author{J\"urgen~K\"onig
\and John~Schliemann
\and Tom\'a\v{s}~Jungwirth
\and Allan~H.~MacDonald}
\authorrunning{J. K\"onig, J. Schliemann, T. Jungwirth, and A.H. MacDonald}
%
%

\maketitle              

\input{allan}

\input{tomas}

\input{jurgen}

\input{john}

\input{concl}

\input{acknow}

\newpage

\input{ref}
%

\end{document}

%% file: allan.tex
\section{Introduction}
\label{section_intro}

Ferromagnetism occurs when Mn is randomly substituted for more than about
2 percent of the cations of several III-V compound semiconductors.
Although only a few host materials have been explored at present, this property
is likely shared by most III-V semiconductors. In this Chapter we will
discuss some of the theoretical pictures that are being developed to explain 
the magnetic and transport  properties of these materials. 
Our development will be based on a phenomenological model that has
been used with great success to explain the sensitivity of bulk and layered
(II,Mn)VI semiconductor optical properties to external magnetic fields.  
(Ferromagnetism does {\em not} occur for Mn in undoped II-VI hosts.)  
The low energy degrees of freedom in this model are holes in the semiconductor 
valence band and one $S=5/2$ local moment for each Mn ion.     

Interest in these ferromagnets was heightened by the demonstration several years
ago that ferromagnetic transition temperatures \cite{ohnosci98} in excess of 
100~K can be achieved in (Ga,Mn)As.  
It has been further heightened recently both by the demonstration of long
spin-coherence times in semiconductors \cite{kikkawanature99} and by the 
dramatic and rapid development of new information storage 
technology based on magnetotransport effects in ferromagnetic 
metals~\cite{metals}. 
It seems clear that semiconductors have many potential advantages over metals for
devices based on the magnetotransport effects that occur in itinerant electron 
ferromagnets, principally because they present a wider canvas for their creative 
manipulation by some combination of impurities, gates and optical excitation.
It is likely that important applications for these materials will be found only 
if ferromagnetism at room temperature can be achieved.  
The very recent discovery \cite{sonoda} of ferromagnetism at temperatures close 
to 1000~K in (Ga,Mn)N has fueled hopes that these materials will indeed 
have technological impact.  Our focus here, however, is on the physics of these ferromagnets; we 
therefore concentrate mainly on the properties of (Ga,Mn)As and (In,Mn)As which have 
been studied most extensively \cite{ohnojmmm99}.  

It is generally accepted that Mn acts as an acceptor when it substitutes for a 
cation in a III-V semiconductor lattice, leaving a Mn$^{2+}$ ion which has a 
half-filled d-shell with angular momentum $L=0$ and spin $S=5/2$.
It is also generally accepted that ferromagnetism occurs 
in these materials because of interactions between Mn local moments that are 
mediated by holes in the semiconductor valence band.  
There is, therefore, a lot of similarity between the ferromagnetism of these 
materials and that of lanthanides and actinides and their compounds in which 
f-electron moments are coupled by d-band itinerant electrons.  
There are also similarities between these materials and the manganite 
compounds that have been extensively studied \cite{manganitereview}
in recent years in part because of the 
large increase in resistance that occurs when $T$ exceeds the Curie temperature, the 
so-called colossal magnetoresistance effect.  
In (III,Mn)V ferromagnets, however, the local moments appear on only a small 
fraction of the atomic sites arranged randomly.
In addition the itinerant electron density is also low, even lower than the Mn local
moment density.  As we explain later this property is likely important in selecting ferromagnetic over 
glassy magnetic order.  The participation of itinerant electrons in the ferromagnetism
of these diluted magnetic semiconductors (DMS) 
adds to their richness, leading in particular to electronic transport
properties that are very sensitive to the magnetic state of the material. 
The physics of ferromagnetic semiconductors is in a sense intermediate between
that of rare earth magnets and that of manganites in that the spin-splitting
of itinerant electron bands due to their exchange coupling with local moments is 
comparable to their Fermi energies, rather than being much smaller than 
band Fermi energies as in the rare earth case or larger as in the 
manganite case.
    
The Chapter is organized as follows.  
In Section~\ref{section_prop} we briefly summarize the main experimental
properties of (III,Mn)V ferromagnets.  
Several different but related approaches that have been explored in an effort to 
gain insight into these materials are outlined in 
Section~\ref{section_approaches}.  
The remaining sections of this review Chapter deal with the development of the 
semi-phenomenological model we favor in which the low energy degrees of freedom 
are exchange-coupled valence-band holes and $S=5/2$ Mn local moments that carry 
a negative charge.
The simplest version of this model is one in which the randomly distributed Mn ions 
are replaced by a uniform continuum, thus completely neglecting disorder.
In Section~\ref{section_mf} we discuss physical predictions based on a 
mean-field treatment of this disorder-free model and demonstrate that it 
successfully describes a number of non-trivial properties of (Ga,Mn)As and 
(In,Mn)As ferromagnets, including their anomalous Hall conductivities.
In Section~\ref{section_sw} we discuss collective excitations of these 
ferromagnets within the disorder-free model, demonstrating that the simple 
mean-field-theory is reasonably reliable for typical parameters of current samples but
must fail at large carrier densities and also in the limit 
of very strong exchange coupling.
In Section~\ref{section_mc}  
 we discuss the results of Monte Carlo calculations that describe the
effect of collective fluctuations of Mn moment orientations. The method
can deal with some of the
complications and additional physics, including the possibility of non-collinear ground
states, that enters when disorder is
added to the theoretical model.
A brief summary is given in Section~\ref{section_end}. 

In Sections~\ref{section_mf}, \ref{section_sw}, and \ref{section_mc}  we have been 
able to present only a small fraction of theoretical data obtained using our approach. An
extensive survey of predicted physical properties of DMS's will be available
at  {\tt http:// unix12.fzu.cz/ ms/index.php} web-pages, launched by the authors
of this Chapter in collaboration with Jan Ku\v{c}era, Byounghak Lee, and Jairo Sinova.

\section{Properties of (III,Mn)V Ferromagnets}
\label{section_prop}

In this section we discuss some important properties of (III,Mn)V ferromagnets 
that have been established by current experiments.  
Thorough recent reviews of the properties of these materials have been prepared 
by Dietl, Matsukura, and Ohno \cite{dietlprb01,handbookarticle}.  
Our objective in this section is to summarize the observations that are most 
important in constraining theoretical descriptions.  
There is at present considerable activity related to the growth and 
characterization of these materials.  We expect rapid progress to be achieved in 
the near future in exploring the range of possible behaviors and relating
them more precisely to molecular-beam-epitaxy growth and post-growth annealing 
protocols.  We list below a number of properties that appear to be safely 
established.  A theoretical picture that is able to explain most of these 
properties is outlined in the following sections.

\begin{itemize}

\item Electron paramagnetic resonance
and optical experiments \cite{szczytkoprb99,linnarssonprb97} 
demonstrate that  $S=5/2$ local moments occur
for dilute concentrations of Mn in GaAs.  These experiments demonstrate that the 
Mn local moment model is correct.

\item The Mn-induced states near the Fermi energy play a key role in the
origin of ferromagnetism and in the magnetotransport properties of
(III,Mn)V DMS's. According to photoemission studies
\cite{okabayashiprb98,okabayashiprb99,okabayashiprb01}, 
those states have As 4$p$ character, i.e.,
can be associated with the host semiconductor valence band states. The angle-resolved
photoemission experiment also showed a negligible shift in the heavy- and light-hole bands
with Mn concentration $x\leq 7$\%.

\item Ferromagnetism is not observed for Mn concentrations  smaller than
$\sim 0.01$ \cite{ohnojmmm99}.  
This property demonstrates that ferromagnetism does not occur when all
valence-band holes are trapped on individual Mn ions or on other defects. 
Antisite defects, for example, are common in  semiconductor samples grown by
low-temperature molecular beam epitaxy (MBE).
For very dilute Mn concentrations, electron spin resonance
experiments demonstrate that most holes are 
trapped not at the Mn acceptors, but at other defects.  

\item The ground-state magnetization, $M(T=0)$, per Mn ion can exceed 
$4 \mu_B$ for larger values of $x$ \cite{ohnojmmm99}.  
Since the magnetization contribution from antiferromagnetically
coupled valence band holes tends to partially compensate the Mn local 
moment magnetization, ferromagnets with these large values of $M(T=0)$ likely have ground states with
(nearly) fully aligned Mn local moments.

\item To date the largest ferromagnetic transition temperatures
in (Ga,Mn)As occur for $x \sim 5$\%.
The current record is $T_c \sim 110 K$ \cite{ohnosci98}.
The drop in critical temperatures at higher $x$ values
may be related to Mn clustering or may have a more fundamental origin.
There is a correlation between large values of $M(T=0)$ and high $T_c$'s.

\item (III,Mn)V thin film ferromagnets grown under compressive 
strain have their magnetic easy axis
in the plane, while ferromagnets grown under tensile
strain have their magnetic easy axis in the growth direction \cite{ohnojmmm99}.  
External magnetic fields $\sim 100$~mT are sufficient to align the magnetization along the
hard axis \cite{ohno96,dietlprb01}. 
These properties can be explained by well understood strain effects 
in the spin-orbit coupled valence bands and demonstrate that the macroscopic properties 
of these ferromagnets are sensitive to details of the valence band electronic structure.

\item The ferromagnetic critical temperature and the temperature dependence of the magnetization 
are altered by post-growth annealing and are sensitive to the details of 
the annealing protocol \cite{ohnojmmm99}.

\item These ferromagnets have large anomalous Hall resistivities \cite{ohnojmmm99}, 
demonstrating
that the itinerant valence bands are full participants in the magnetism.
The large anomalous Hall resistivities reflect the strong spin-orbit 
coupling that is present at the top of the valence band in zincblende semiconductors.    

\item The semiconductor valence bands are spin-split in the ferromagnetic state.  
These  semiconductor ferromagnets exhibit strong magnetoresistance effects, 
like tunnel magnetoresistance \cite{hayashijcg99}, 
that are characteristic of itinerant electron ferromagnets, again 
demonstrating that the itinerant electrons are full participants in the magnetism.
\end{itemize}

\section{Theoretical Approaches} 
\label{section_approaches}

Ferromagnetism is a collective effect due to interactions between electrons.  If it were
possible to do so, we would explain its microscopic origins in a particular class of materials 
by solving the many-electron Schr\"{o}dinger equation directly, given the position 
of all the nuclei. Fortunately this uninteresting direct approach is impossible, now and forever more,
because of the macroscopic number of interacting electronic degrees of freedom.
Instead we must resort to some combination of approximation and phenomenological modeling,
settling on the correct approach only after careful comparison with experiment.  This is 
the art of condensed matter science; an intricate tango between theory and 
experiment leading to a conclusion that cannot be anticipated while the dance is in progress.   
In some cases, fractional quantum Hall systems and possibly cuprate superconductors
for example, the low-energy degrees of freedom in terms of which observable physical
properties are best described are not simply related to the bare electrons which 
appear in the many-electron Hamiltonian.  Often, however, the low-energy degrees of freedom
are more obvious.

A practical approach to many-electron physics that is often 
successful is spin-density-functional (SDF) theory, in which many-body effects appear in 
exchange-correlation potential contributions to effective independent-particle
Hamiltonians.  SDF theory has the advantage that it is a first principles approach 
without any phenomenological parameters.   Among its many achievements is a generally satisfactory
description of itinerant electron ferromagnetism in transition metals.
SDF theory has been applied \cite{bscalcs}
to (III,Mn)V ferromagnetism by Sanvito {\it et al.} and by 
van Schilfgaarde and Mryasov.  To date these calculations have been performed 
using the local density approximation (LDA) of SDF theory, an approximation that is 
not reliable when local moments are formed, i.e., when strong correlations
suppress fluctuations in the number of electrons in the d-shell or the f-shell of a 
particular atom.  LDA-SDF theory supercell \cite{bscalcs} calculations predict that  
majority spin d-electrons of a Mn atom substituted on a cation site of GaAs
lie at the Fermi energy, rather than lying well below the Fermi energy as they
would if the half-filled d-shell formed a $S=5/2$ local moment.   
Similar results have been obtained in coherent potential approximation band-structure
calculations \cite{akaiprl98} for (In,Mn)As, and are clearly in disagreement with 
experiment in both cases.  In local moment systems, it is necessary to account for the 
increase in instantaneous cite energy when the occupancy of a localized orbital 
is increased.  The LDA+U method has been developed to mitigate this deficiency of SDF theory  
and has recently\cite{parkphysb00} been applied to (III,Mn)V ferromagnets and finds 
a dominant As 4$p$ orbital weight at the Fermi energy, consistent with the photoemisson experiment.

\begin{figure}
\centerline{\includegraphics[width=7.8cm]{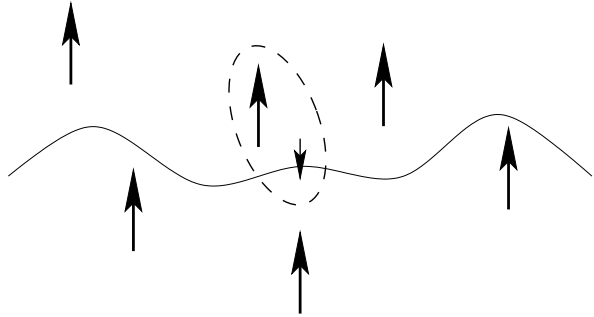}}
\caption{Model for (III,Mn)V semiconductors: local magnetic moments (Mn$^{2+}$)
  with spin $S=5/2$ are antiferromagnetically coupled to itinerant carriers 
  (holes) with spin $s=1/2$.}
\label{fig_model}
\end{figure}

It appears likely that a lot of detailed information on the electronic properties of 
(III,Mn)V ferromagnetic semiconductors can be reliably obtained from LDA+U SDF calculations and 
we expect this approach will play an important role in the modeling of these materials in the 
future.  Our work, however, follows a semi-phenomenological strategy,
starting from a model in which the local-moment character of the Mn d-orbitals is 
asserted rather than derived.  These models are 
adapted from ones used \cite{kossutpss76,furdynasemi88} 
to describe the optical properties of 
(II,Mn)VI semiconductors.  The low-energy degrees of freedom in the 
{\em kinetic-exchange} model \cite{larsonprb88,dietlhandbook94}
we employ are $S=5/2$ local moments representing half-filled
Mn d-shells, and holes in the Mn valence band (see Fig.~\ref{fig_model}).
Since Mn$^{2+}$ ions should act as acceptors when substituted on the cation 
sites of III-V semiconductors, we would expect one hole per Mn if no other 
charged defects were present in the system.  
However, antisite defects are common when III-V semiconductor films are grown by 
MBE at the low temperatures required to prevent Mn segregation. 
The hole density and the Mn density are therefore taken as 
separate sample-dependent quantities, to be determined experimentally.  
The Hamiltonian of this model is specified in detail in the following paragraph. 
There has also been theoretical work on these materials
based on a still simpler model \cite{bhattrefs} where holes are assumed to hop only between 
Mn acceptor sites, where they interact with the Mn moments via phenomenological
exchange interactions.  These models have some advantages in getting at the physics of the 
dilute Mn limit, and can also easily be adapted to include the holes that 
are localized on ionized antisite defects rather than Mn acceptors
\cite{chudnovskiy}.

Like any phenomenological model, the one we use is defined most fundamentally by its 
low-energy degrees of freedom.  Also important, however, is the Hamiltonian that 
acts in the implied Hilbert space.  The length scales associated with holes in 
these compounds are still long enough that a $\vec k \cdot \vec p$, envelope function,
description \cite{laserbook} 
of the semiconductor valence bands is appropriate and we take that approach
here.  The operators in terms of which the phenomenological Hamiltonian is expressed 
include the spin operator $\vec S_{I}$ for the $S=5/2$ local moment on site $I$ and
the multi-band envelope function hole spin density operator $\vec s(\vec r)$.
The following key terms are included in the minimal version of the model Hamiltonian:
\begin{itemize} 

\item[a)] The coupling of the Mn spin to the external magnetic field,
$g \mu_B \sum_I \vec S_{I} \cdot \vec H_{\rm ext}$.

\item[b)] The band Hamiltonian of the host III-V semiconductor, usually
described using a multi-band envelope function formalism 
\cite{laserbook,abolfathprb01}.
For many properties it is necessary to incorporate spin-orbit coupling in a 
realistic way.  Six- or eight-band models that include the `split-off' band 
and/or the conduction band are sometimes desirable.  
Unlike the local moment models, the Hilbert space includes all host lattice 
sites for each hole rather than only localized orbitals centered on the Mn sites.
This band Hamiltonian should include the strain effects due to lattice matching between 
the epitaxially grown (III,Mn)V films and the substrate on which they are grown.

\item[c)] Antiferromagnetic exchange coupling between the Mn$^{2+}$ spin and 
valence-band holes, $J_{\rm pd} \sum_I \vec S_{I} \cdot \vec s(\vec R_{I})$.  
This interaction represents virtual coupling to states that have been 
{\em integrated out} of the model's Hilbert space, ones in which electrons are 
exchanged between the Mn ion d shells and the valence 
band \cite{larsonprb88,dietlhandbook94}.
The exchange interactions are isotropic to a good approximation because the 
Mn$^{2+}$ ion has total angular momentum $L=0$. 
Experimental estimates for $J_{\rm pd}$ vary from $150\pm 40$~meV nm$^3$ 
\cite{ohnosci98}, to $68\pm 10$~meV nm$^3$ \cite{omiyaphyse}, to 
$55\pm 10$~meV nm$^3$ \cite{okabayashiprb98}.  Recent experimental work limits $J_{pd}$ 
to a value toward the lower end of this range and fixes its value within perhaps 
$20\%$. 

\end{itemize}

The terms d)--f), listed below, are necessary to describe the crossover to the 
localized limit in which holes are bound either to Mn acceptors or to other defects.  
Note that simpler, impurity-band models assume that the system is in this limit 
from the outset, much as our model assumes from the outset that the Mn d-shells 
form local moments.  There are sometimes technical difficulties in describing 
the localized limit with our higher-level model, so that considerable 
simplification arises from using the impurity-band model.
There is however, a penalty to pay since the model does not apply to the regimes of 
greatest interest in which the band electrons are not localized.  Even
when impurity models do apply, it is difficult to guess at appropriate distribution functions for
the inter-site hopping parameters that play a key role.    
\begin{itemize} 

\item[d)] The attractive Coulomb interaction between the ionized Mn$^{2+}$ 
acceptor and a valence-band hole.  
In an envelope function formalism, {\em central-cell} corrections to the 
interaction are necessary to capture the isolated bound-acceptor limit
accurately \cite{Bhatta:00}.  

\item[e)] The repulsive Coulomb interaction among holes.  
This interaction is key in screening the ionized Mn$^{2+}$ acceptors and cannot 
be neglected except in the completely-localized-hole limit.  
When it is included, it usually must be approximated in a way which avoids
artificial hole-hole self-interactions.

\item[f)] The repulsive interaction between holes and ionized antisite (group-V 
element on group-III site) defects.  The antisite defects compensate for 
the Mn acceptors and reduce the overall hole density, in addition to providing an 
important additional scattering center.  In the dilute Mn limit, experiment \cite{fedorych}  
suggests that most Mn$^{2+}$ ions do not have bound holes, possibly due to this
compensation.  When all holes are strongly localized, most Mn local moments will
be free and the system will not have ferromagnetic order.

\end{itemize}

The following terms  are potentially important in some circumstances.
\begin{itemize}

\item[g)] The scalar scattering potential that represents the energy difference 
between a valence-band p electron on a host site and a valence-band p electron on a 
Mn site.  This effect has normally been excluded 
since its size and sign is not yet know.

\item[h)] Direct exchange interactions between Mn ions on neighboring sites.  
These terms result from microscopic processes in which exchange of electrons between 
the valence band and two nearby Mn d-shells is correlated.  Terms of this type are known
to be important in (II,Mn)VI semiconductors, but appear to be less important in
(III,Mn)V semiconductors.

\item[i)] Direct coupling of band electrons to external magnetic field.
\end{itemize}

In the rest of this Chapter we discuss only pictures of (III,Mn)V ferromagnetism 
that follow from the minimal model that includes only the a)--c) Hamiltonian terms.

%% file: tomas.tex
\section{Mean-field-theory predictions}
\label{section_mf}

In this and the following section we make an important approximation that 
achieves a drastic simplification.
We will refer to this as the {\em continuum Mn approximation}, although it has 
sometimes been referred to as a virtual crystal approximation.  
It is motivated by the observation that the Fermi wavelength of 
the valence-band electrons is typically longer than the distance between Mn ions,
mainly because the Mn acceptors are compensated and also partially because there 
are four occupied valence bands.  
When the Mn ion distribution is replaced by a continuum with the same 
spin-density, randomness is completely eliminated from the minimal model.  
This approximation has many elements in common with the dynamic mean-field-theory
(DMFT) approximation, applied \cite{millis} to these ferromagnets recently by 
Chattopadhyay {\em et al.}, although the DMFT does retain some of the 
consequences of randomness neglected here and in the following section.  
The possibility of starting with a description based on an 
approximation where the random Mn distribution is replaced by a continuum
emphasizes an essential difference between ferromagnetic semiconductors and 
classical spin glass systems in which dilute Mn local moments are distributed randomly 
in a metallic host.  In both cases it is true that the interaction between
local moments mediated by the itinerant electrons is oscillatory and ferromagnetic for 
separations smaller than the itinerant electron Fermi wavelength.  However, the Fermi wavelength
is longer than the distance between local moments in the doped semiconductor case,
whereas it is shorter in the spin glass case \cite{abrikosov}.     
Each Mn ion interacts ferromagnetically with several of its neighbors.
The continuum Mn approximation will fail in the limit of dilute Mn ions and 
also if the exchange interaction between band and Mn spins is too strong.
This approximation does not allow for the sensitivity of magnetic properties to 
annealing protocols that has been established in experiment.  It does, however,  
seem to be reliable in the limit of principle 
interest, that of high Mn densities and high critical temperatures,
where the holes are metallic and their interaction with Mn acceptors will be 
effectively screened.  The merit of this approximation is that it enables quantitative 
prediction of many physical properties.  In this section we discuss three important 
properties, the ferromagnetic transition temperature, the magnetic anisotropy energy, and 
the anomalous Hall conductance.
This work described below is
motivated by the view in science, it is ultimately up to experiment to decide on the
reliability of any approximation made in modeling a physical system.  As we will point 
out, the utility of the continuum Mn approximation is strongly supported by observations.
In fact, it is not a surprise  that this approximation is a good starting point, 
given its success in describing the influence of external fields on the 
properties of the closely related paramagnetic (II,Mn)VI semiconductors
\cite{furdynasemi88,dietlhandbook94}.
The present section makes in addition a {\em mean-field}
approximation by ignoring correlations between Mn and band spin configurations.

Our mean-field theory is derived in the spin-density-functional framework
and leads to a set of physically transparent coupled equations
\cite{jungwirthprb99}. The effective magnetic field seen by localized
magnetic ions consists of an external magnetic field and the 
mean kinetic-exchange-coupling contribution from spin-polarized carriers,
\begin{equation}
{\vec H}_{\rm eff}({\vec R}_I) = {\vec H}_{\rm ext} +
         J_{\rm pd} \langle {\vec s}({\vec R}_I) \rangle /  g \mu_B \; ,
\label{Heff}
\end{equation}
where 
$\langle {\vec s}({\vec R}_I) \rangle$ is the carrier spin density at Mn sites,
and $g$ is the g-factor of the local moments.
The mean spin polarization of a  magnetic ion is given by \cite{aharoni}
\begin{equation}
\langle {\vec S} \rangle_I = -
        S B_{S}\big(S g \mu_B H_{\rm eff}({\vec R}_I) / k_B T\big)
        {\hat H_{\rm eff}}({\vec R}_I)\; ,
\label{SI}
\end{equation}
where $B_S(x)$ is the Brillouin function and ${\hat H_{\rm eff}}({\vec R}_I)$
is the unit vector along the direction of the effective magnetic field
defined in Eq.~(\ref{Heff}).
The itinerant-hole spin density is determined by solving the
Schr\"{o}dinger equation for holes which experience 
a kinetic-exchange effective Zeeman field ${\vec h}(\vec r)$.  
The field ${\vec h}(\vec r)$ is non-zero only in the ferromagnetic state
and,
in the continuum limit, reads 
\begin{equation}
{\vec h}(\vec r) = J_{\rm pd} \  N_{\rm Mn}(\vec r)\ \langle 
{\vec S} \rangle(\vec r)\; ,
\label{h}
\end{equation}
where $N_{Mn}=4x/a_{lc}^3$ is the Mn density in Mn$_{x}$III$_{1-x}$V zincblende
semiconductors with a lattice constant $a_{lc}$.
For inhomogeneous systems such as quantum wells or
superlattices, the itinerant holes experience also an 
electrostatic potential due to heterostructure confinement
and external bias (if present). Using the local-spin-density
approximation (LSDA), itinerant hole-hole interaction can be
accounted for by including an additional spin-dependent one-particle
potential in the Schr\"{o}dinger equation.

\subsection{Ferromagnetic transition temperature}

In homogeneous DMS systems, the hole-spin density $\langle s \rangle$ and the 
kinetic-exchange potential $h$ are related at small $h$ by
\begin{equation}
  \langle s \rangle= -\frac{\chi_f}{(g^{\ast} \mu_B)^2}h\; .
\label{bulk}
\end{equation}
Here, $g^{\ast}$ is the hole g-factor and
$\chi_f$ is the interacting hole magnetic susceptibility,
\begin{equation}
  \frac{\chi_f}{(g^{\ast}\mu_B)^2}=-\frac{d^2(E_{\rm tot}/V)}{dh^2}\; ,
\label{chi}
\end{equation}
where $E_{\rm tot}/V$ is the total energy density of the itinerant-hole
system.
The Curie-Weiss transition temperature, obtained from Eqs.~(\ref{Heff})
- (\ref{bulk}) in the $H_{\rm ext}=0$ limit, is 
\begin{equation}
  k_B T_c = \frac{N_{\rm Mn} S (S+1)}{3} 
  \frac{J_{\rm pd}^2\chi_f}{(g^{\ast} \mu_B)^2} \; .
\label{tc}
\end{equation}

To understand the qualitative physics implicit in this $T_c$-equation~(\ref{tc}),
we discuss first the magnetic susceptibility expressions of a model itinerant electron 
system with a single spin-split band and an effective mass $m^{\ast}$.
The kinetic-energy contribution $E_{\rm tot}^{\rm kin}$ to the total energy gives
\begin{equation}
  \frac{d^2(E_{\rm tot}^{\rm kin}/V)}{dh^2}=-\frac{m^{\ast}k_F}{4\pi^2\hbar^2}
  \; ,
\label{etk}
\end{equation}
where $k_F$ is the Fermi wavevector. The exchange energy of the
spin-polarized parabolic-band model adds a contribution 
\begin{equation}
  \frac{d^2(E_{\rm tot}^{\rm exch}/V)}{dh^2}=
  -\frac{e^2 (m^{\ast})^2}{4\pi^3\varepsilon \hbar^4} \; ,
\label{etx}
\end{equation}
where $\varepsilon$ is the dielectric constant of the host semiconductor.
At high hole densities $p$, the kinetic-energy term dominates and $T_c$
is proportional to the Fermi wavevector, i.e., to $p^{1/3}$. 
Equations~(\ref{etk}) and (\ref{etx})
also show that the band contribution to the
mean-field $T_c$ increases linearly with $m^{\ast}$
while the exchange enhancement of $T_c$ is proportional to
$(m^{\ast})^2$.
Note that correlation effects, not discussed here in detail, suppress 
the mean-field $T_c$
by only $\sim 1\%$ for typical 
experimental hole densities ($p\sim 0.1$~nm$^{-3}$) in 
bulk (III,Mn)As ferromagnets.

\begin{figure}
\centerline{\includegraphics[width=8cm]{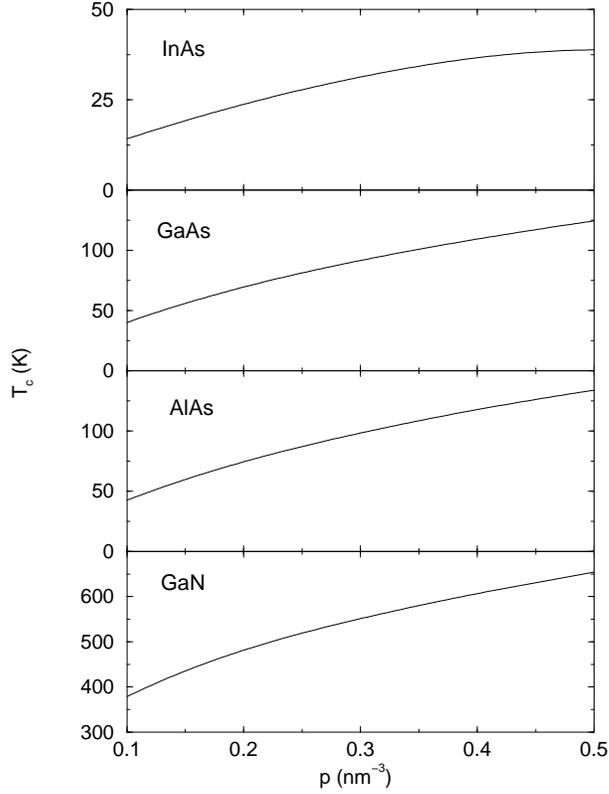}}
\caption{The band (kinetic energy) 
contribution to the mean-field ferromagnetic critical temperature $T_c$ 
for Mn concentration $x=5$\% 
is plotted as a function of
hole density $p$ for InAs, GaAs, AlAs, and GaN host semiconductors.
}
\label{tcmf}
\end{figure}

To obtain quantitative predictions for
$T_c$, it is necessary to evaluate the kinetic and exchange contribution to the itinerant hole
susceptibility using a realistic six-band Kohn-Luttinger model \cite{abolfathprb01},
instead of the parabolic band model.  The Hamiltonian
contains the spin-orbit splitting parameter $\Delta_{\rm so}$ and
three other phenomenological parameters, $\gamma_1$, $\gamma_2$, and $\gamma_3$,
whose values for the specific III-V host can be found, e.g., 
in Refs.~\cite{dietlprb01,iiiv}.
Results \cite{dietlprb01,dietlsci00,jungwirthphyse01,jungwirthtbp} are plotted 
in Fig.~\ref{tcmf} as a function of the hole density for InAs, GaAs, AlAs, and 
GaN host semiconductors.  
In the density range considered, only the two heavy-hole and two light-hole 
bands are occupied in the arsenides.  However, the mixing
between these four bands and the two spin-orbit split-off bands is
strong and must be accounted for. In GaN, spin-orbit
coupling is weak and all six bands are occupied by holes.
The numerical data  are consistent with the qualitative 
analysis based on the parabolic band model: the band (kinetic-energy) 
contribution to $T_c$ follows roughly the $p^{1/3}$ dependence, the exchange
enhancement of $\sim 10$\% is only weakly density dependent. 
The $T_c$ values at a given density are ordered according to
the heavy-hole and light-hole masses in the arsenide hosts.
For GaN, with all six bands occupied, the simple model of a parabolic spin-split
band is less instructive. Yet the large numerical $T_c$'s in this material are 
consistent with the large heavy-hole mass, nearly twice as large as in 
AlAs.  

The mean-field prediction for the critical temperature agrees quantitatively 
with the experimental value of 110~K measured in Mn-doped GaAs with 
Mn concentration $x=5$\% and $p=0.35$~nm$^{-3}$.
Thermal fluctuations neglected by the mean-field theory,
discussed in the following section, reduce the 
theoretical $T_c$ estimate by less than 5\% \cite{jungwirthtbp}, explaining the 
quantitative success of the mean-field theory in this sample.  The same analysis
finds approximately a $T_c$ suppression \cite{jungwirthtbp} of approximately
20\% compared to mean-field theory due for (Ga,Mn)N, implying that room temperature 
ferromagnetism may occur in III-V DMS.

\subsection{Magnetic anisotropy}

Experiments \cite{ohno96,munekataapl93} in (III,Mn)V DMS's have demonstrated
that these ferromagnets have remarkably square hysteresis loops and that the 
magnetic easy axis is dependent on epitaxial growth lattice-matching strains.
The physical origin \cite{dietlprb01,abolfathprb01} of the anisotropy energy in 
our model is spin-orbit coupling in the valence band. 
Even in mean-field theory, we find that the magnetic anisotropy physics of
these materials is rich and that easy axis reorientations can occur as a 
function of sample parameters including hole density or epitaxial growth 
lattice-matching strains. 

Magnetic anisotropy in the absence of strain is well described by a
cubic harmonic expansion truncated at sixth order, an approximation commonly
used in the literature \cite{skomskicoey} on magnetic materials.  The
corresponding cubic harmonic expansion for total energy of a system
of non-interacting holes in the presence of the effective field 
$h$ is
\begin{equation}
  {E_{\rm tot}(\hat M) \over V} = {E_{\rm tot}(\langle100\rangle) \over V} 
  + K^{\rm ca}_1 ({\hat h}_x^2  {\hat h}_y^2 + {\hat h}_y^2 {\hat h}_z^2 
  + {\hat h}_x^2 {\hat h}_z^2)
  + K^{\rm ca}_2 \; {\hat h}_x^2 {\hat h}_y^2 {\hat h}_z^2 \; ,
\label{gammaofm}
\end{equation}
where $\hat h$ is the unit vector along the  field $h$.
The cubic anisotropy coefficients $K_{1}^{\rm ca}$ and $K_{2}^{\rm ca}$ are 
related to total energies for $\hat{h}$ along the high symmetry crystal 
directions by following expressions:
\begin{eqnarray}
  K_{1}^{\rm ca} &=& \frac{4 [E_{\rm tot}(\langle110\rangle) - E_{\rm tot}
    (\langle100\rangle)]}{V}
  \nonumber \\ 
  K_{2}^{\rm ca} &=& \frac{27 E_{\rm tot}(\langle111\rangle) - 36
    E_{\rm tot}(\langle110\rangle) + 9 E_{\rm tot} (\langle100\rangle)}{V}. 
  \label{cubicanisotropy}
\end{eqnarray}

MBE growth techniques produce (III,Mn)V films whose lattices are locked to 
those of their substrates.  
X-ray diffraction studies \cite{ohnojmmm99} have established that
the resulting strains are not relaxed by dislocations or other defects, even
for thick films. Strains in the (III,Mn)V film break
the cubic symmetry assumed in Eq.~(\ref{gammaofm}). However, the influence
of MBE growth lattice-matching strains on the hole bands of cubic
semiconductors is well understood \cite{laserbook} and we can use the
same formal mean-field theory as in the previous subsection
to account for strain effects on
magnetic anisotropy.

\begin{figure}
\centerline{\includegraphics[width=8cm]{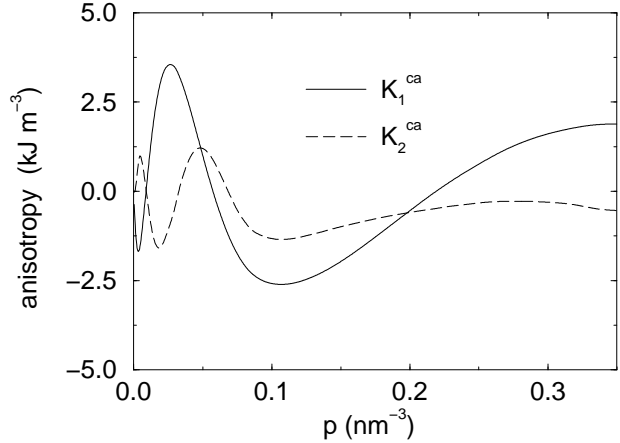}}
\caption{Cubic magnetic anisotropy coefficients $K_1^{\rm ca}$ and $K_2^{\rm ca}$
as a function of hole density $p$.}
\label{aniso}
\end{figure}

We turn now to a series of illustrative calculations intended to closely
model the ground state of (Ga,Mn)As. For  Mn
density $N_{\rm Mn}=1$~nm$^{-3}$ ($x \approx 5$\%),
$h \approx 140$~meV at zero temperature.  
This value of $h$ is not so much
smaller than the spin-orbit splitting parameter in GaAs \cite{iiiv,dietlprb01}
($\Delta_{\rm so}=341$~meV), so that accurate calculations require the six-band
Luttinger model \cite{abolfathprb01}.   
Even with $N_{\rm Mn}$ fixed, our calculations show that the magnetic
anisotropy of (III,Mn)V ferromagnets is strongly
dependent on both hole density and strain. The hole density can be varied by
changing growth conditions or by adding other dopants to the material, and
strain in a (Ga,Mn)As film can be altered by changing
substrates. The cubic anisotropy coefficients (in units
of energy per volume) for strain-free material are plotted as a function of
hole density in  Fig.~\ref{aniso}. 
The easy axis is nearly always determined by the leading cubic anisotropy
coefficient $K_1^{\rm ca}$, except near values of $p$ where this coefficient
vanishes. As a consequence, the easy axis in strain free samples is almost always 
either along one of the cube edge directions ($K_1^{\rm ca} > 0$), or along one 
of the cube diagonal directions ($K_1^{\rm ca} < 0$). Transitions in which the
easy axis moves between these two directions occur twice over the range of hole
densities studied.  (Similar transitions occur as a function of $h$, and
therefore temperature, for fixed hole density.) Near the hole density
$p= 0.01$~nm$^{-3}$, both anisotropy coefficients nearly vanish and a fine-tuned 
nearly perfect isotropy is achieved. The slopes of the anisotropy coefficient curves vary
as the number of occupied bands increases from $1$ to $4$ with
increasing hole density. This behavior is clearly seen from the correlation
between oscillations of the anisotropy coefficients
and onsets of higher band occupations.

\begin{figure}
\centerline{\includegraphics[width=8cm]{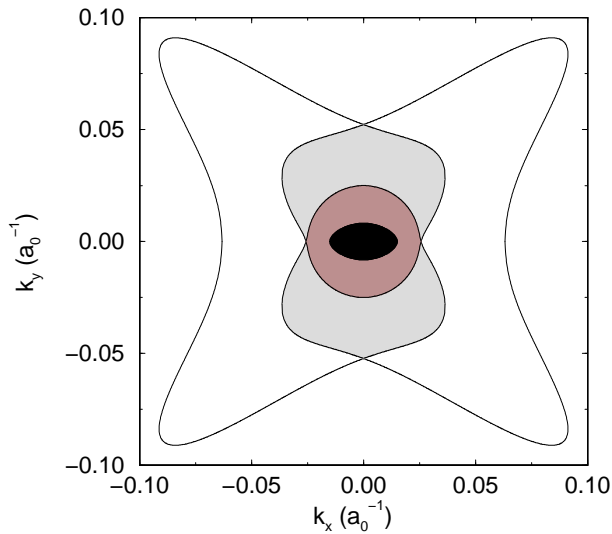}}
\caption{Six-band model
Fermi surface intersections with the $k_z=0$ plane for $p=0.1~{\rm nm}^{-3}$  
and $h=140$~meV.  This figure is for magnetization orientation along   
the $\langle100\rangle$ direction.}
\label{nfl_6_100}
\end{figure}

\begin{figure}
\centerline{\includegraphics[width=8cm]{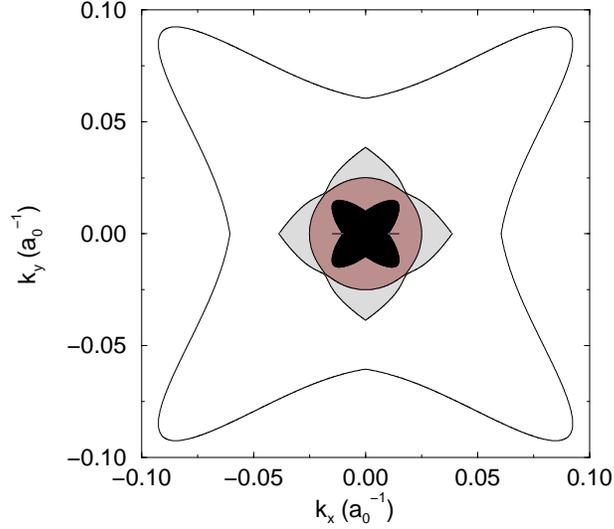}}
\caption{Six-band model
Fermi surface intersections with the $k_z=0$ plane for the parameters of
Fig.~\protect\ref{nfl_6_100} and magnetization orientation along the
$\langle110\rangle$ direction.}
\label{nfl_6_110}
\end{figure}

Six-band model Fermi surfaces are illustrated in
Figs.~\ref{nfl_6_100} and \ref{nfl_6_110} 
by plotting their intersections with the
$k_z=0$ plane at $p =0.1~{\rm nm}^{-3}$ for the cases of $\langle100\rangle$
and $\langle110\rangle$ ordered moment orientations.
The dependence of quasiparticle band
structure on ordered moment orientation, apparent in comparing these figures,
should lead to large anisotropic magnetoresistance effects in (Ga,Mn)As
ferromagnets.  We also note that in the case of
cube edge orientations, the Fermi surfaces of different bands intersect.  This
property could have important implications for the decay of long-wavelength
collective modes.

\begin{figure}
\centerline{\includegraphics[width=8cm]{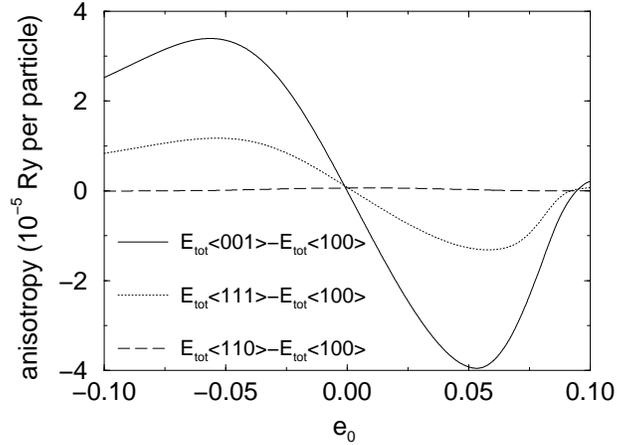}}
\caption{Energy differences among $\langle001\rangle$,
$\langle100\rangle$, $\langle110\rangle$, and $\langle111\rangle$ magnetization
orientations vs. in-plane strain $e_0$ at $h=140$~meV and $p = 0.35$~nm$^{-3}$.
For compressive strains ($e_0<0$), the system has an easy
magnetic plane perpendicular to the growth direction. For  tensile strains
($e_0>0$), the anisotropy is easy-axis with the preferred magnetization
orientation along the growth direction. The anisotropy changes sign at
large  tensile strain.}
\label{anisostrain}
\end{figure}

In Fig.~\ref{anisostrain} we present mean-field theory predictions for the
strain-dependence of the anisotropy energy at $h=140$~meV and hole
density $p=0.35~{\rm nm}^{-3}$. According to our calculations, the easy axes in
the absence of strain are along the cube edges in this case. 
The relevant value of 
the in-plane strain produced by the substrate-film lattice mismatch,
\begin{equation}
e_0=\frac{a_s-a_f}{a_f},
\end{equation}
depends on the substrate on which the epitaxial (Ga,Mn)As
film is grown. The most
important conclusion from Fig.~\ref{anisostrain} is that strains as small
as $1\%$ are sufficient to completely alter the magnetic anisotropy energy
landscape.  For example for (Ga,Mn)As on
GaAs, $e_0=-0.0028$ at $x=0.05$.  The
anisotropy has a relatively strong uniaxial contribution, even for this
relatively modest compressive strain, which favors in-plane moment
orientations \cite{dietlprb01,abolfathprb01}, 
in agreement with experiment \cite{ohnojmmm99}. 
A relatively small ($\sim$1 kJ
m$^{-3}$) residual in-plane anisotropy remains which favors $\langle110\rangle$
over $\langle100\rangle$.  For $x=0.05$ (Ga,Mn)As on a $x=0.15$ (In,Ga)As
buffer the strain is tensile, $e_0 =0.0077$, and we predict a substantial
uniaxial contribution to the anisotropy energy which favors growth direction
orientations \cite{dietlprb01,abolfathprb01}, 
again in agreement with experiment \cite{ohnojmmm99}.  For the
tensile case, the anisotropy energy changes more dramatically than for
compressive strains due to the depopulation of higher subbands. 
At large tensile strains, the sign of the anisotropy changes,
emphasizing the subtlety of these effects and the latitude which exists for
strain-engineering of magnetic properties.

\subsection{Anomalous Hall effect}
The mean-field description of hole bands in the presence of exchange 
coupling to the localized Mn moments provides a starting point for building
a theory of transport in (III,Mn)V ferromagnets. Here we concentrate on
the anomalous Hall effect which is an important sample characterization
tool in ferromagnetic systems.
The Hall resistivity of ferromagnets has an ordinary contribution,
proportional to the external magnetic-field strength, and an anomalous
contribution usually assumed to be proportional
to the sample magnetization. In our approach \cite{jungwniumacdtbp}, 
the anomalous Hall conductance of a homogeneous ferromagnet is related 
to the Berry phase of the electronic wavefunction acquired by
a cyclic evolution along the Fermi surface.

In the standard model of the AHE in metals, 
skew-scattering \cite{smitphysica58} and side-jump \cite{bergerprb70} scattering
give rise to contributions to the Hall resistivity 
proportional to the diagonal resistivity $\rho$ 
and $\rho^2$ respectively, with the latter process tending to 
dominate in alloys because $\rho$ is larger.
Our evaluation of the AHE in (III,Mn)V
ferromagnets is based on a theory \cite{sundaramprb99} of semiclassical
wave-packet dynamics 
which 
implies a contribution to the Hall conductivity that is 
independent of the kinetic-equation scattering term.  The interest in this 
contribution is motivated in part by practical considerations, since our
current understanding of (III,Mn)V ferromagnets is not sufficient
to permit confident modeling of quasiparticle scattering.  
The relation of our approach to standard theory is reminiscent 
of disagreements between Smit \cite{smitphysica58} and 
Luttinger \cite{luttingerpr58} 
that occurred early in the development of AHE theory and do not appear 
to have ever been fully resolved.  We follow
Luttinger \cite{luttingerpr58} in taking the view that there is a contribution
to the AHE due to the change in wavepacket group velocity that occurs
when an electric field is applied to a ferromagnet.  
The electron group velocity correction is conveniently evaluated using 
expressions derived by Sundaram and Niu \cite{sundaramprb99}:
\begin{equation}
\dot {\vec x}_{c} = \frac{\partial \epsilon}{\hbar \partial \vec k}
+ (e/\hbar) \vec E \times \vec \Omega.
\label{velocity}
\end{equation}
The first term on the right-hand-side of Eq.~(\ref{velocity}) is 
the standard Bloch band group velocity.  Our anomalous Hall conductivity 
is due to the second term, proportional to the Berry curvature
$\vec{\Omega}$, defined below.  
It follows from symmetry considerations that for a cubic semiconductor 
under lattice-matching strains and with $\hat m$ aligned by
external fields along the $\langle 001\rangle$ growth direction, only the 
$z$-component of $\vec \Omega$ is nonzero: 
\begin{equation}
\Omega_z(n,\vec k) = 2\, {\rm Im}
 \big[ \langle \frac{\partial u_n}{\partial k_y} \vert \frac{\partial 
u_n}{\partial k_x} \rangle 
\big].
\label{omega}
\end{equation}
Here $|u_n\rangle$ is the periodic part of the $n$-th Bloch band
wavefunction with the mean-field spin-splitting term included in the Hamiltonian.
The anomalous Hall conductivity that results from this velocity correction is
\begin{equation}
\sigma_{\rm AH} = -\frac{e^2}{\hbar} \sum_{n} \int \frac{d \vec k}{(2 \pi)^3} 
 f_{n,\vec k} \Omega_{z}(n,\vec k)\; ,
\label{sigmaH}
\end{equation}
where $f_{n,\vec k}$ is the equilibrium Fermi occupation factor for the band
quasiparticles.  We have taken the convention that a positive $\sigma_{\rm AH}$
means that the anomalous Hall current is in the same direction as the normal 
Hall current.  

This Berry phase contribution to the anomalous Hall conductance
occurs in any itinerant electron ferromagnet with spin-orbit coupling. 
To assess its importance for (III,Mn)V ferromagnets we first explore a simplified model that yields  
parabolic dispersions for the two heavy--hole and two light--hole bands, and  
neglect coupling to the split-off band by assuming a large
spin-orbit coupling \cite{laserbook,abolfathprb01}. 
Detailed numerical simulations accounting for the mixing of the spin-orbit 
split-off bands and warping of the occupied heavy--hole and light--hole 
bands \cite{laserbook,abolfathprb01} in the (In,Mn)As and (Ga,Mn)As 
samples \cite{ohnoprl92,ohnosci98} will follow.
Within the 4-band spherical model, the spin operator $\vec s=\vec j/3$,
and the Hamiltonian for holes in III-V host semiconductors
can be written as

\begin{equation}
H_0= {\hbar^2 \over 2 m} \left[
  (\gamma_1+\frac52\gamma_2)k^2-2\gamma_2(\vec k\cdot\vec j)^2 \right]
\; ,
\label{h0}
\end{equation}
where $\vec j$ is the total angular momentum operator and
$\gamma_1$ and $\gamma_2$ are the Luttinger parameters \cite{laserbook,iiiv}.
In the unpolarized case ($h=0$), the total Hamiltonian, $H=H_0-hj_z/3$ (the external
magnetic field is assumed to be in the $+\hat z$ direction), is diagonalized 
by spinors $|j_{\hat k}\rangle$ where, e.g.,  
$j_{\hat k}\equiv \vec j\cdot\hat k=\pm3/2$ for the two degenerate heavy--hole
bands with the effective mass $m_{\rm hh} = m/(\gamma_1-2\gamma_2)$.      
The corresponding Berry phase, 
$\int d^2k \Omega(\pm3/2,\vec k)=\pm3/2(\cos\theta_{\vec k}-1)$,
is largest at the equator ($\cos\theta_{\vec k}\equiv k_z/k_{\rm hh}=0$) 
and vanishes at the poles ($|\cos\theta_{\vec k}|=1$) of the spherical Fermi 
surface of radius $k_{\rm hh}$. Because of the band degeneracy,
the anomalous Hall conductivity (\ref{sigmaH}) vanishes in the $h=0$ limit.
The effective Zeeman coupling present in the 
ferromagnetic state both modifies the Fermi surface shapes and  
renormalizes the Berry phases.   
Up to linear order in $h$ we obtain that 
$k_{\rm hh}^{\pm}=k_{\rm hh}\pm hm_{\rm hh}/(2\hbar^2 k_{\rm hh})\,
\cos\theta_{\vec k}$ and that the Berry phase is reduced (enhanced) by a factor
$[1\mp 2mh/(9\gamma_2\hbar^2 k_{\rm hh}^2)]$. 
A similar analysis for the light-hole bands leads to the total net contribution 
to the AHE from the four bands whose lower and upper bounds are:
\begin{equation}
  \frac{e^2}{4\pi^2\hbar^3}h(3\pi^2p)^{-1/3}m_{\rm hh}
  <\sigma_{\rm AH}<
  \frac{e^2}{4\pi^2\hbar^3}h(3\pi^2p)^{-1/3}2^{2/3}
  m_{\rm hh} \;.
\label{sigmaH4b}
\end{equation}
Here $p=k_{\rm hh}^3/3\pi^2\, (1+\sqrt{m_{\rm lh}/m_{\rm hh}})$ is the total 
hole density and $m_{\rm lh} = m/(\gamma_1+2\gamma_2)$ is the light-hole
effective mass. The lower bound in Eq.~(\ref{sigmaH4b}) is obtained assuming 
$m_{\rm lh} \ll m_{\rm hh}$ while the upper bound is reached when 
$m_{\rm lh}\approx m_{\rm hh}$.

\begin{figure}
\centerline{\includegraphics[width=8cm]{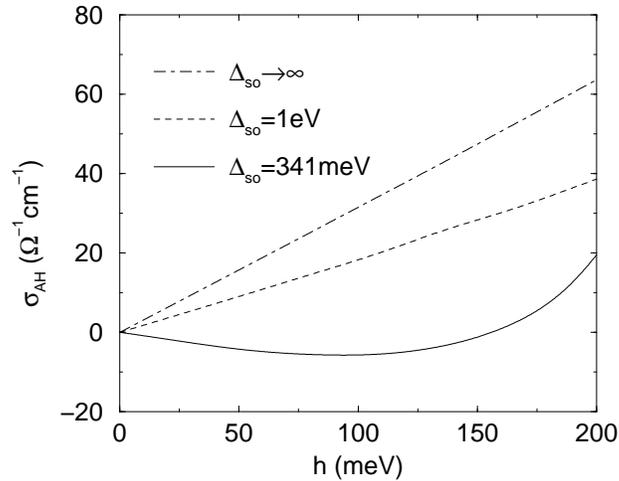}}
\caption{
Illustrative calculations of the
anomalous Hall conductance as a function of the band-splitting 
effective Zeeman field for hole
density $p=0.35$~nm$^{-1}$. The dotted-dashed curve was obtained
assuming infinitely large spin-orbit
coupling and the decrease of theoretical $\sigma_{\rm AH}$ with decreasing
spin-orbit coupling strength is demonstrated for $\Delta_{\rm so}=1$~eV 
(dashed line) and $\Delta_{\rm so}=341$~meV (solid line).
}
\label{fignature1}
\end{figure}

Based on the above analysis we draw the following conclusions: The anomalous
velocity due to the Berry phase can have a sizable effect on the AHE in (III,Mn)V
ferromagnets. The solid line in Fig.~\ref{fignature1} shows our analytic results
for the GaAs effective masses $m_{\rm hh}=0.5m_e$ and $m_{\rm lh}=0.08m_e$. Note that 
in experiment, anomalous Hall conductances are in order of 1--10 
${\rm \Omega^{-1}\, cm^{-1}}$ and the  effective exchange field 
$h\sim 10-100$~meV.  A large $\sigma_{\rm AH}$ is expected in systems with large 
heavy--hole effective mass and with the ratio $m_{\rm lh}/m_{\rm hh}$ close to 
unity. 

\begin{figure}
\centerline{\includegraphics[width=8cm]{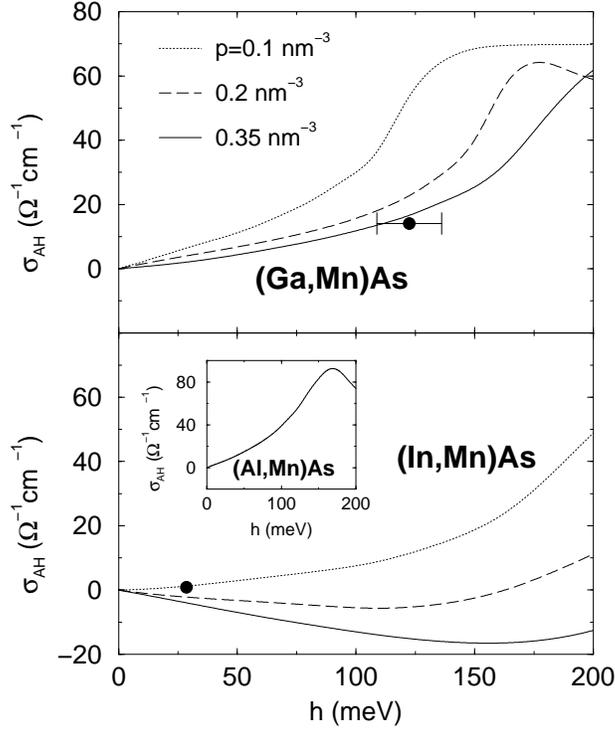}}
\caption{  
Full numerical simulations of $\sigma_{\rm AH}$ for GaAs host (top panel), InAs 
host (bottom panel), and AlAs host (inset) with hole densities 
$p=0.1$~nm$^{-1}$ (dotted lines), $p=0.2$~nm$^{-1}$ (dashed lines),
and $p=0.35$~nm$^{-1}$ (solid lines).
The filled circles in the top and bottom panels represent measured 
AHE \protect\cite{ohnoprl92,ohnojmmm99} values.
The saturation mean-field $h$ values for the two points were estimated from 
nominal sample parameters \protect\cite{ohnoprl92,ohnojmmm99}. Horizontal error 
bars correspond to the experimental uncertainty of the $J_{\rm pd}$ coupling 
constant. The measured hole density in the (Ga,Mn)As sample is  
$p=0.35$~nm$^{-1}$; for (In,Mn)As, $p=0.1$~nm$^{-1}$ was determined
indirectly from the sample's transition temperature. 
}
\label{fignature2}
\end{figure}

So far we have discussed the limits of infinitely strong spin-orbit coupling and 
weak effective exchange field, relative to the hole Fermi energy. In the opposite
limits of zero spin-orbit coupling or large $h$, $\sigma_{\rm AH}$ vanishes. 
This implies that the anomalous Hall conductivity is generally nonlinear in the 
exchange field or the magnetization.  
To explore the intermediate regime we diagonalized the six-band 
Luttinger Hamiltonian numerically \cite{laserbook,abolfathprb01} with the spin-orbit gap
$\Delta_{\rm so}=1$~eV as well as for the GaAs value $\Delta_{\rm so}=341$~meV. 
Results shown in Fig.~\ref{fignature1} confirm that a smaller $\sigma_{\rm AH}$ is 
expected in systems with smaller $\Delta_{\rm so}$ and suggest that both 
positive and negative signs of $\sigma_{\rm AH}$ can occur, in general. The 
curves in Fig.~\ref{fignature1} are obtained by neglecting band warping in 
III-V semiconductor compounds. The fact that valence bands in these materials
are typically strongly non-parabolic, even in the absence of the field $h$ and in
the large $\Delta_{\rm so}$ limit, is accurately captured by introducing the 
third phenomenological Luttinger parameter 
$\gamma_3$ \cite{laserbook,abolfathprb01}. 
Numerical data including all Luttinger parameters
indicates that warping tends to lead to an increase of $\sigma_{\rm AH}$, as seen when 
comparing solid curves in Fig.~\ref{fignature1} and in the top panel of 
Fig.~\ref{fignature2}.
The hole-density dependence of $\sigma_{\rm AH}$, illustrated in 
Fig.~\ref{fignature2}, is qualitatively consistent with the spherical model 
prediction (\ref{sigmaH4b}).
Also in accord with the outlined chemical trends, numerical data in 
Fig.~\ref{fignature2} suggest large positive AHE in (Al,Mn)As, intermediate 
positive $\sigma_{\rm AH}$ in (Ga,Mn)As, and a relatively weaker AHE in 
(In,Mn)As with the sign of $\sigma_{\rm AH}$ that may depend on the detail
structure of the sample. 

We make now a comparison between our $\sigma_{\rm AH}$ calculations and 
experimental data in the (In,Mn)As and (Ga,Mn)As samples, analyzed in detail by 
Ohno and coworkers \cite{ohnoprl92,ohnosci98,ohnojmmm99}. The nominal Mn 
densities in the two measured systems are $N_{\rm Mn}=0.23$~nm$^{-3}$ for
the InAs host and $N_{\rm Mn}=1.1$~nm$^{-3}$ for the GaAs host, yielding 
saturation values of the effective field $h=25\pm3$~meV and $h=122\pm14$~meV,
respectively.
The low-temperature hole density of the (Ga,Mn)As sample, $p=0.35$~nm$^{-3}$, 
was unambiguously determined \cite{ohnojmmm99} from the ordinary Hall 
coefficient measured at high magnetic fields of 22-27~T. 
Since a similar experiment has not been reported for the (In,Mn)As sample we 
estimated the hole density, $p=0.1$~nm$^{-3}$, by matching the measured 
ferromagnetic transition temperature $T_c=7.5$~K to the density dependent 
mean-field $T_c$. 

As demonstrated in Fig.~\ref{fignature2}, our theory explains the order of
magnitude difference between experimental AHE in the two samples 
($\sigma_{\rm AH} \approx 1\,{\rm \Omega}^{-1}$~cm$^{-1}$ in (In,Mn)As and 
$\sigma_{\rm AH} \approx 14\,{\rm \Omega}^{-1}$~cm$^{-1}$ in (Ga,Mn)As). 
The calculations are also consistent with the observed positive sign and 
monotonic dependence of $\sigma_{\rm AH}$ on sample 
magnetizations \cite{ohnojmmm99}.

We take the  agreement in both magnitude and sign of the AHE 
as a strong indication that the anomalous
velocity contribution dominates the AHE in  homogeneous (III,Mn)V ferromagnets.
This Berry phase term, which is independent of quasiparticle
scatterers,  is relatively easily evaluated with high accuracy,
enhancing the utility of the Hall measurement in sample characterization.
The success of this model  also supports the use of the 
simple mean-field approximation discussed in this section, in which Mn ions
are represented by a uniform density continuum, 
to describe at least the ground state of these ferromagnets.

%% file: jurgen.tex
\section{Collective excitations within a continuum picture}
\label{section_sw}

\subsection{Beyond mean-field theory and RKKY interaction}

The power and the success of the mean-field picture employed in the previous 
section lies in the fact that it is a simple theoretical approach which makes 
it easy to calculate many observable quantities numerically.
Mean-field theory, however, neglects correlation between local-moment spin
configurations and the free-carrier state and, therefore, fails to describe 
the existence of low-energy long-wavelength spin excitations, among other things.
Because of its neglect of collective magnetization fluctuations, mean-field 
theory, e.g., always overestimates the ferromagnetic critical temperature.
There are many examples in itinerant electron systems where mean-field
theory overestimates ferromagnetic transition temperatures by more than an order
of magnitude and it is not {\em a priori} obvious that mean-field theory will
be successful in (III,Mn)V ferromagnets.  Indeed, we will find that the 
multi-band character of the semiconductor valence band plays an essential role
in enabling high ferromagnetic transition temperatures in these materials.

In this section we identify the elementary spin excitations, determine their
dispersion, and discuss implications for the Curie temperature 
\cite{koenigprl00,koenigspringer01,schliemannapl01,schliemanncm33,koenigcm16}.
The starting point of our analysis is the itinerant-carrier-mediated 
ferromagnetic interaction between local magnetic moments.
Such an interaction is provided by the familiar Ruderman-Kittel-Kasuya-Yoshida 
(RKKY) theory.
The RKKY picture, however, only applies as long as the perturbation induced 
by the Mn spins on the itinerant carriers is small.
As we will derive below, the proper condition is $\Delta \ll \epsilon_F$ where
$\Delta = N_{\rm Mn} J_{\rm pd} S$ is the (zero-temperature) spin-splitting gap 
of the itinerant carriers due to an average effective field induced by the Mn 
ions, and $\epsilon_F$ is the Fermi energy.
This condition is, however, never satisfied in (III,Mn)V ferromagnets,
partially because the valence-band carrier concentration $p$ is usually much 
smaller than the Mn impurity density $N_{\rm Mn}$.  
A related drawback of the RKKY picture is that it assumes an instantaneous
static interaction between the magnetic ions, i.e., the dynamics of the free 
carriers are neglected.
We will see below that this dynamics is important to obtain all types of 
elementary spin excitations.
As a consequence, {\em RKKY theory does not provide a proper description 
of the ordered state in ferromagnetic DMSs.}

As in the previous section, we use here the minimal model including terms 
a)--c) of Section~\ref{section_approaches} and employ the Mn continuum 
approximation.
Extensions to the minimal model may be important in some circumstances.
They are, however, not essential for the general discussion in the
present section, which will attempt to explain the considerations that 
determine when collective effects neglected by mean-field-theory are important. 

\subsection{Independent spin-wave theory for parabolic bands}

The main idea of our theory is to derive an effective description for the Mn 
spin system by integrating out the valence-band carriers and to look for 
fluctuations of the Mn spins around their spontaneous mean-field magnetization
direction (which we choose as the $z$-axis).
Using the Holstein-Primakoff (HP) representation \cite{auerbach94}, we express 
the Mn spins in terms of bosonic degrees of freedom.
We expand the effective action up to quadratic order, i.e., we treat the spin 
excitations as noninteracting Bose particles.
From the corresponding propagator we deduce the dispersion of all elementary
spin excitations.

To keep the discussion transparent we start with a two-band model for the 
itinerant carriers with quadratic dispersion.
Later, in Section~\ref{subsec:realistic}, we extend our theory to a model 
with a more realistic band structure described by a six-band Kohn-Luttinger 
Hamiltonian.

For small fluctuations around the mean-field magnetization, we can write the 
spin operators as
\begin{eqnarray}
   S^+ ({\vec r}) &\approx& b({\vec r}) \sqrt{2N_{\rm Mn}S}
\\
   S^- ({\vec r}) &\approx& b^\dag({\vec r}) \sqrt{2N_{\rm Mn}S}
\\
   S^z ({\vec r}) &=& N_{\rm Mn}S - b^{\dag}({\vec r}) b({\vec r})
\label{eq:sz}
\end{eqnarray}
with bosonic fields $b^{\dag}({\vec r}), b({\vec r})$.
The state with fully polarized Mn spins (along the $z$-direction) corresponds, 
in the HP boson language, to the vacuum with no bosons.
The creation of a HP boson reduces the magnetic quantum number by one.

The partition function $Z$ can be expressed as a coherent-state path integral 
in imaginary time over the HP bosons and the valence-band carriers, which are
fermions.
Since the Hamiltonian is bilinear in the fermionic fields, we can integrate out
the itinerant carriers and arrive at an effective description in terms of the 
impurity spin degree of freedom labeled by the complex number coherent state labels
for the boson fields, $z$ and $\bar z$.
We get $Z = \int {\cal D} [\bar z z] \exp (- S_{\rm eff} [\bar z z])$ with the 
effective action
\begin{equation}
  S_{\rm eff} [\bar z z] =  S_{\rm BP} [\bar z z]
  - \ln \det \left[ (G^{\rm MF})^{-1} + \delta G^{-1}(\bar zz) \right] \, ,
\label{effective action}
\end{equation}
where $S_{\rm BP} [\bar z z] = \int_0^\beta d\tau \int d^3 r \, \bar z 
\partial_\tau z$ is the usual Berry's phase term.
In Eq.~(\ref{effective action}), we have already split the total kernel 
$G^{-1}$ into a mean-field part $(G^{\rm MF})^{-1}$ and a fluctuating part 
$\delta G^{-1}$,
\begin{eqnarray}
   (G^{\rm MF})^{-1}_{ij} &=&
 \left( \partial_\tau -\mu \right)\delta_{ij} + \langle i| H_0 | j \rangle
        + N_{\rm Mn}J_{\rm pd}S s^z_{ij}
\label{kernel MF}
\\
   \delta G^{-1}_{ij}(\bar zz) &=& {J_{\rm pd}\over 2} \left[ \left( 
        z s^-_{ij} + \bar z s^+_{ij} \right) \sqrt{2N_{\rm Mn}S} 
        - 2 \bar z z s^z_{ij} \right]
\label{kernel fluc}
\end{eqnarray}
where $\mu$ denotes the chemical potential, and $i$ and $j$ range over a 
complete set of hole-band states (i.e., here, for the model with two parabolic
bands, $i$ and $j$ label band wavevectors and spin, $\uparrow$,$\downarrow$), and 
$s^{z}_{ij}$ and $s^{\pm}_{ij}$ are matrix elements of the itinerant-carrier 
spin matrices.
The combination $\Delta = N_{\rm Mn}J_{\rm pd}S$ defines the mean-field energy 
to flip the spin of an itinerant carrier.
The physics of the itinerant carriers is embedded in the effective action
of the magnetic ions.
It is responsible for the retarded and non-local character of the interactions 
between magnetic ions.

So far we have made no approximations.  
The independent spin-wave theory is obtained by expanding 
Eq.~(\ref{effective action}) up to quadratic order in $z$ and $\bar z$, i.e., 
spin excitations are treated as noninteracting HP bosons.
This is a good approximation at low temperatures, where the number of spin
excitations per Mn site is small.

We obtain (in the imaginary time Matsubara and coordinate Fourier 
representation) an action that is the sum of the temperature-dependent 
mean-field contribution and a fluctuation action.
The latter is
\begin{equation}
\label{quad action}
  S_{\rm eff} [\bar z z] = {1\over \beta V} \hspace{-1mm}
  \sum_{|\vec k| \le k_D, m} \! \!
  \bar z(\vec k, \nu_m) D^{-1}(\vec k, \nu_m) z(\vec k, \nu_m).
\end{equation}
A Debye cutoff $k_D$ with $k_D^3 = 6 \pi^2 N_{\rm Mn}$ ensures that we include 
the correct number of magnetic-ion degrees of freedom, $|\vec k| \le k_D$.
The kernel of the quadratic action defines the inverse of the spin-wave 
propagator,
\begin{equation}
   D^{-1}(\vec k, \nu_m) = 
        - i \nu_m 
        + {J_{\rm pd} p\xi \over 2} 
        + {N_{\rm Mn} J_{\rm pd}^2 S \over 2 V} \sum_{{\vec q}} 
          { f [\epsilon_\uparrow({\vec q})] 
          - f [\epsilon_\downarrow({\vec q}+{\vec k})] 
          \over 
          i\nu_m + \epsilon_\uparrow({\vec q}) 
                 - \epsilon_\downarrow({\vec q}+{\vec k}) }
\,\,\,
\label{inverse propagator}
\end{equation}
where $\xi=(p_\downarrow - p_\uparrow)/p$ is the fractional free-carrier spin 
polarization, and $\epsilon_{\uparrow,\downarrow}({\vec q})$ is the energy of
spin-up and spin-down valence-band holes, 
$\epsilon_{\uparrow,\downarrow}({\vec q}) = \epsilon_q \pm \Delta/2$, and
$\epsilon_q = \hbar^2 q^2 /(2m^*)$.
The second term of Eq.~(\ref{inverse propagator}) is the the energy for
a Mn spin excitation in mean-field-theory, $\Omega^{\rm MF} =  J_{\rm pd} p\xi / 2 = x \Delta$. 
It differs from the itinerant-carrier spin splitting by the ratio of the 
spin densities $x=p \xi /(2N_{\rm Mn}S)$, which is always much smaller than 1 
in (III,Mn)V ferromagnets.
Mean-field theory is, thus, recovered by dropping the last term in 
Eq.~(\ref{inverse propagator}).
It is this term that describes the response of the free-carrier system to
changes in the magnetic-ion configuration.

\subsection{Elementary spin excitations}

We obtain the spectral density of the spin-fluctuation propagator by
analytical continuation, $i\nu_m \rightarrow \Omega+i0^+$ and 
$A(\vec k,\Omega)={\rm Im} \, D(\vec k,\Omega)/\pi$.
In the following we consider the case of zero temperature, $T=0$.
We find three different types of spin excitations \cite{koenigprl00}.

\subsubsection{Goldstone-mode spin waves.}

\begin{figure}
\centerline{\includegraphics[width=8cm]{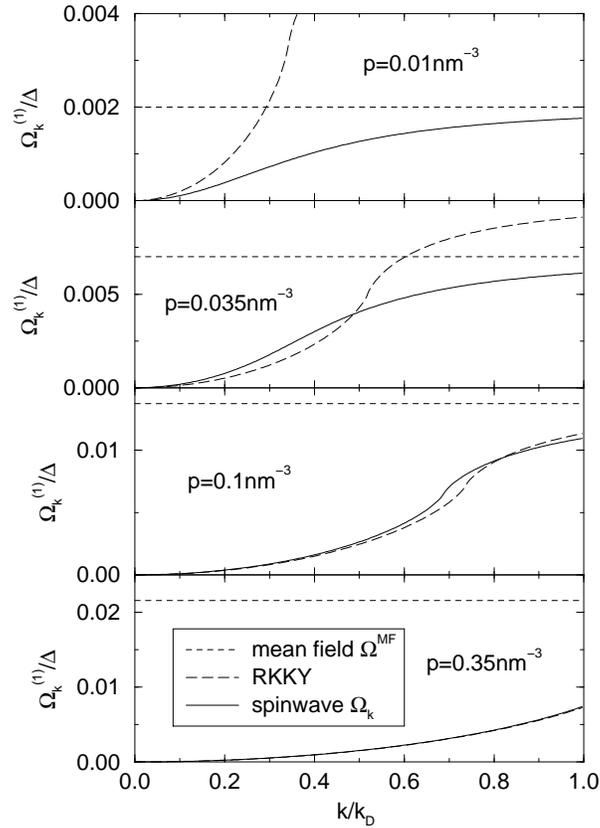}}
\caption{Spin-wave dispersion (solid lines) for $J_{\rm pd}=0.06{\rm eVnm}^3$, 
        $m^*=0.5m_e$, $N_{\rm Mn}=1{\rm nm}^{-3}$, and four different 
        itinerant-carrier concentrations $p = 0.01 \, {\rm nm}^{-3}$, 
        $0.035 \, {\rm nm}^{-3}$, $0.1 \, {\rm nm}^{-3}$, and 
        $0.35 \, {\rm nm}^{-3}$.
        The ratio $\Delta/\epsilon_F$ is $2.79$, $1.21$, $0.67$, and $0.35$, 
        which yields the fractional free-carrier spin polarization $\xi$ as
        $1$, $1$, $0.69$, and $0.31$.
        The short wavelength limit is the mean-field result 
        $\Omega^{\rm MF}=x\Delta$ (short-dashed lines), and the long-dashed
        lines are the result obtained from an RKKY picture.}
\label{dispersion1}
\end{figure}

Our model has a gapless Goldstone-mode branch reflecting the spontaneous 
breaking of spin-rotational symmetry.
The dispersion of this low-energy mode for four different valence-band carrier 
concentrations $p$ is shown in Fig.~\ref{dispersion1} (solid lines).
At large momenta, $k\rightarrow \infty$, the spin-wave energy approaches
the mean-field result $\Omega^{(1)}_{k} \rightarrow \Omega^{\rm MF}$ 
(short-dashed lines in Fig.~\ref{dispersion1}).
Expansion of the $T=0$ propagator for small momenta yields for the 
collective modes dispersion,
\begin{equation}
\label{eq:sw_general}
   \Omega^{(1)}_k = { x/\xi\over 1- x} \epsilon_k \left( {3+2\xi\over 5} -
 {4 \over 5} \xi {\epsilon_F\over \Delta} \right) + {\mathcal O}(k^4) \, ,
\end{equation}
where $\epsilon_F$ is the Fermi energy of the majority-spin band.
In strong and weak-coupling limits, $\Delta \gg \epsilon_F$ and  
$\Delta \ll \epsilon_F$, respectively, Eq.~(\ref{eq:sw_general}) simplifies to
\begin{eqnarray}
\label{eq:sw_strong}
   \Omega^{(1)}_k = { x \over 1- x} \epsilon_k  + {\mathcal O}(k^4) 
        \hspace*{2.38cm} {\rm for} \qquad \Delta \gg \epsilon_F\, , 
\\
\label{eq:sw_weak}
   \Omega^{(1)}_k = { p\over 32N_{\rm Mn}S} \epsilon_k \left( 
        \Delta \over \epsilon_F \right)^2 + {\mathcal O}(k^4) 
        \qquad {\rm for} \qquad  \Delta \ll \epsilon_F\, . 
\end{eqnarray}
We note that the dependence of the spin-wave energy on the system parameters,
namely the exchange interaction strength $J_{\rm pd}$, hole concentration $p$, 
local-impurity density $N_{\rm Mn}$, and effective mass $m^*$ is different in 
these two limits, indicating that the microscopic character of the gapless 
collective excitations differs qualitatively in the two limits. 
The energy of long-wavelength spin waves is determined by a competition
between exchange and kinetic energies.
To understand this in more detail one can impose the spin configuration of a 
static spin wave on the Mn spin system, evaluate the ground-state energy of the 
itinerant-carrier system in the presence of the generated exchange field, and 
compare this with the ground-state energy of a uniformly polarized state.
The results of this calculation are explained briefly below;
for details see Ref.~\cite{koenigspringer01}. 
Given the Mn spin configuration, the valence-band carriers can 
either follow the spatial dependence of the Mn spin density in order to 
minimize the exchange energy, as they do in the strong-coupling limit $\Delta \gg \epsilon_F$,
or minimize the kinetic energy by forming a state with a homogeneous spin polarization
, as they do in the weak-coupling limit $\Delta \ll \epsilon_F$.
The corresponding energy scales are provided by $\Delta$ and $\epsilon_F$, 
i.e., the crossover from one regime to the other is governed by the ratio 
$\Delta/\epsilon_F$.

\subsubsection{Stoner continuum.}

We observe that the frequency $\nu_m$ is not only present in the first term
of Eq.~(\ref{inverse propagator}), it enters the third term, too.
This is the reason why, in addition to the Goldstone mode, other spin 
excitations can appear in our model.
They are absent in a static-limit description, i.e., when the frequency 
dependence of the third term of Eq.~(\ref{inverse propagator}) is neglected.

\begin{figure}
\centerline{\includegraphics[width=8cm]{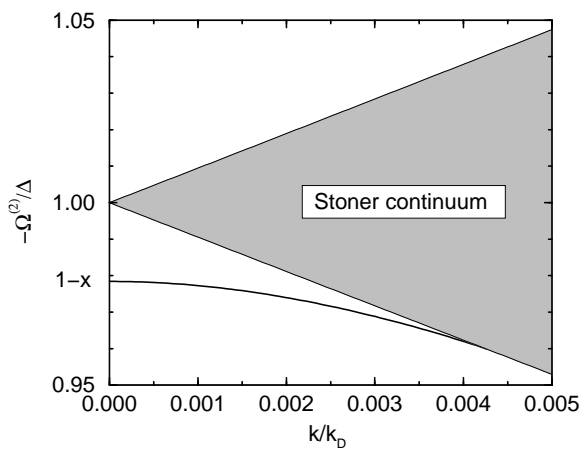}}
\caption{Stoner excitations and optical spin-wave mode in the free-carrier 
        system for $J_{\rm pd}=0.06{\rm eVnm}^3$, $m^*=0.5m_e$, 
        $N_{\rm Mn}=1{\rm nm}^{-3}$, and $p=0.35{\rm nm}^{-3}$.
        In an RKKY picture these modes are absent.}
\label{dispersion2}
\end{figure}
We find a continuum of Stoner spin-flip particle-hole excitations.
They correspond to flipping a single spin in the itinerant-carrier system and,
since $x \ll 1$, occur in this simple model at much larger energies near the itinerant-carrier 
spin-splitting gap $\Delta$ (see Fig.~\ref{dispersion2}).
For $\Delta > \epsilon_F$ and zero temperature, all these excitations carry 
spin $S^z=+1$, i.e., increase the spin polarization.  
They therefore turn up at negative frequencies in the boson propagator we 
study.
When $\Delta < \epsilon_F$, excitations with both $S^z=+1$ and $S^z=-1$ 
contribute to the spectral function.
The continuum lies between the curves $-\Delta-\epsilon_k\pm 
2\sqrt{\epsilon_k\epsilon_{F}}$ and for $\Delta < \epsilon_F$ also between
$-\Delta+\epsilon_k\pm 2\sqrt{\epsilon_k(\epsilon_F-\Delta)}$.

\subsubsection{Optical spin waves.}

We find additional collective modes analogous to the optical spin waves
in a ferrimagnet.
Their dispersion lies below the Stoner continuum (see Fig.~\ref{dispersion2}).
At small momenta the dispersion is
\begin{equation}
   -\Omega^{(2)}_{k} = \Delta (1-x) - {\epsilon_k \over 1-x}
        \left( {4\epsilon_F \over 5x\Delta} - {2-(2-5x)/\xi \over 5x}\right)
        + {\cal O}(k^4) \, .
\end{equation}
The finite spectral weight at negative frequencies indicates that, because of
quantum fluctuations, the ground state is not fully spin polarized.

\subsection{Comparison to RKKY and to the mean-field picture}

For comparison we evaluate the $T=0$ magnon dispersion assuming an RKKY 
interaction between magnetic ions.  
This approximation results from our theory if we neglect spin polarization 
in the itinerant carriers and evaluate the static limit of the resulting 
spin-wave propagator defined in Eq.~(\ref{inverse propagator}).
The Stoner excitations and optical spin waves shown in Fig.~\ref{dispersion2} 
are then not present and the Goldstone-mode dispersion is incorrect except when
$\Delta\ll\epsilon_F$, as depicted in Fig.~\ref{dispersion1} (long-dashed 
lines).

In the mean-field picture, correlations among the Mn spins are neglected. 
The mean-field theory can be obtained in our approach by taking the Ising 
limit, i.e., replacing ${\vec S}\cdot{\vec s}$ by $S^z s^z$.
As mentioned before, this amounts to dropping the last term in 
Eq.~(\ref{inverse propagator}).
The energy of an impurity-spin excitation is then dispersionless, 
$\Omega^{\rm MF}=x\Delta$ (short-dashed line in Fig.~\ref{dispersion1}), and 
always larger than the real spin-wave energy.

\subsection{Spin-wave dispersion for realistic bands}
\label{subsec:realistic}

For a quantitative analysis \cite{koenigcm16} of the spin-wave dispersion 
we extend our parabolic-band model to a six-band Kohn Luttinger Hamiltonian.  
The effective action for the HP bosons describing the Mn impurity spins is
given by the same formal expression Eq.~(\ref{effective action}) with the 
contributions Eqs.~(\ref{kernel MF}) and (\ref{kernel fluc}) to the kernel.
The difference is that for each Bloch wavevector 
$i$ and $j$ now label the states in a six-dimensional
Hilbert space (instead of two dimensions for spin up and down), $H_0$ is the
Kohn-Luttinger Hamiltonian, and $s^z_{ij}$ and $s^\pm_{ij}$ are $6\times 6$ 
matrices.

The next step is again an expansion of the effective action up to quadratic
order in $z$ and $\bar z$.
In the two-band model, where spin is a good quantum number, only 
$\bar z({\vec k},\nu_m) z({\vec k},\nu_m)$ the combinations appear, see 
Eq.~(\ref{quad action}).  Since the coherent state labels can 
can be viewed as boson creation and annihilation operators, these contributions are 
diagonal in total boson number. 
In the presence of spin-orbit coupling, however, spin is no longer a good 
quantum number, and the combinations 
$\bar z({\vec k},\nu_m) \bar z(-{\vec k},- \nu_m)$ and 
$z({\vec k},\nu_m) z(-{\vec k},- \nu_m)$ which increase or 
decrease the number of HP bosons, come into play.

Since our aim here is to derive the dispersion relations of the low-energy spin
waves, rather than to address the full excitation spectrum including the 
Stoner continuum and the optical spin waves, we take the  
static limit as discussed in the context of the two-band model.
After a Bogoliubov transformation, we obtain for the spin-wave energy
\begin{equation}
  {\Omega_{\vec k}\over \Delta} = { J_{\rm pd} \over 2}
        \sqrt{ \left( { p\xi \over \Delta }
        - E_{\vec k}^{+-} \right)^2
        - \left|E_{\vec k}^{++}\right|^2 } 
\label{dispersion}
\end{equation}
with the definition
\begin{equation}
        E_{\vec k}^{\sigma \sigma'}  = - {1\over V} \sum_{\vec q} 
        \sum_{\alpha\beta}
        { f [\epsilon_\alpha({\vec q})] - 
          f [\epsilon_\beta({\vec q}+{\vec k})] \over
          \epsilon_\alpha({\vec q}) - \epsilon_\beta ({\vec q}+{\vec k}) } 
        s^\sigma_{\alpha\beta} s^{\sigma'}_{\beta\alpha}
\label{E}
\end{equation}
for $\sigma,\sigma'=\pm$.
The indices $\alpha$ and $\beta$ label the single-particle eigenstates for
valence-band carriers at a given wavevector ${\vec q}$ and ${\vec q}+{\vec k}$,
and $s^\pm_{\alpha\beta} = \langle \alpha |s^\pm | \beta \rangle$.
The remaining task is to evaluate the fractional itinerant-carrier 
polarization $\xi$ and the quantities $E_{\vec k}^{+-}$ and $E_{\vec k}^{++}$
numerically.

\begin{figure}
\centerline{\includegraphics[width=8.cm]{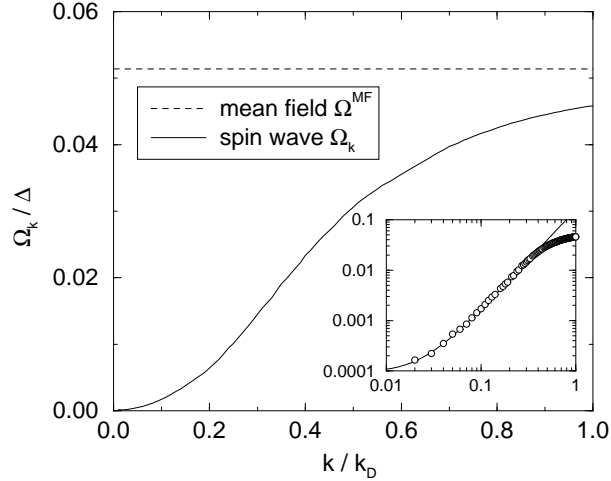}}
\caption{Main panel: Spin-wave dispersion for the 6-band model for 
        itinerant-carrier density $p=0.35 \, {\rm nm}^{-3}$, impurity-spin 
        concentration $N_{\rm Mn} = 1.0 \, {\rm nm}^{-3}$ and exchange 
        coupling $J_{\rm pd} = 0.068 \, {\rm eV \, nm}^{-3}$.
        Inset: Spin-wave dispersion on a log-log plot (circles) and the 
        parabolic fit (solid line).}
\label{dispersion_six_band}
\end{figure}
In Fig.~\ref{dispersion_six_band} we show the spin-wave dispersion for 
wavevectors ${\vec k}$ along the easy axis obtained using parameters valid for 
(Ga,Mn)As \cite{ohnojmmm99}.
We observe that the effect of $E_{\vec k}^{++}$ in Eq.~(\ref{dispersion}) is
negligibly small and can, therefore, be dropped.  
Furthermore, we find that the dispersion is fairly independent of its 
wavevector direction, a property that is usually implicitly assumed in 
micromagnetic descriptions of magnetic materials.

\subsubsection{Spin stiffness.}

The quantized energy of a long-wavelength spin wave in a ferromagnet with
uniaxial anisotropy can be written as
\begin{equation}
   \Omega_{k} = {2K\over N_{\rm Mn}S} 
        +  {2A\over N_{\rm Mn}S} \, k^2 + {\cal O}(k^4) \, ,
\label{gap+stiffness}
\end{equation}
where $K$ is the anisotropy energy constant, and $A$ denotes the spin stiffness
or exchange constant.
While the anisotropy constant can be obtained from the mean-field energy for
different magnetization orientations (see previous section), the virtue of the 
spin-wave calculation is to extract the spin stiffness as well.

\begin{figure}
\centerline{\includegraphics[width=8.cm]{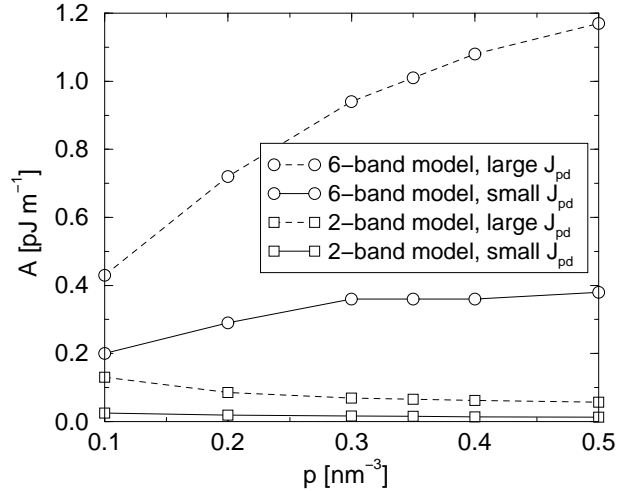}}
\caption{Exchange constant $A$ as a function of itinerant-carrier density $p$
        for the six-band and the two-band model for two different values of
        $J_{\rm pd} = 0.068 \, {\rm eV \, nm}^{-3}$ (solid lines) and 
        $0.136 \, {\rm eV \, nm}^{-3}$ (dashed lines).
        The impurity-spin concentration is chosen as 
        $N_{\rm Mn} = 1.0 \, {\rm nm}^{-3}$, which yields
        $\Delta = 0.17 \, {\rm eV}$ (solid lines) and 
        $\Delta = 0.34 \, {\rm eV}$ (dashed lines), respectively.}
\label{fig_stiffness}
\end{figure}

In Fig.~\ref{fig_stiffness} we show the spin stiffness $A$ as a function of 
the itinerant-carrier density for two values of $J_{\rm pd}$ for both the 
isotropic two-band and the full six-band model.
We find that the spin stiffness is much larger for the six-band calculation 
than for the two-band model.
Furthermore, for the chosen range of itinerant-carrier densities the trend is 
different: in the two-band model the exchange constant decreases with 
increasing density, while for the six-band description we observe an increase
with a subsequent saturation.

To understand this behavior we recall that the two-band model predicts a 
different dependence of $A$ on $p$ in the strong and weak-coupling limits with 
a crossover near $\Delta \sim \epsilon_F$, see Eqs.~(\ref{eq:sw_strong}) and
(\ref{eq:sw_weak}).
The difference in the trends seen for the two- and six-band model in 
Fig.~\ref{fig_stiffness} is explained in part by the observation that, at 
given itinerant-carrier concentration $p$, the Fermi energy $\epsilon_F$ is 
much smaller when the six-band model is employed, where more bands are 
available for the carriers, than in the two-band case.
Furthermore, we emphasize that, even in the limit of low carrier concentration,
it is not only the (heavy-hole) mass of the lowest band which is important for 
the spin stiffness.
Instead, a collective state in which the spins of the itinerant carriers 
follow the spatial variation of a Mn spin-wave configuration will involve the 
light-hole band, too.  Our calculations show that accounting for 
the presence of this second more dispersive band is essential to understanding 
the success of mean-field theory.  {\em Crudely, the large mass heavy
hole band dominates the spin-susceptibility and enables local magnetic
order at high temperatures, while the dispersive light hole band
dominates the spin stiffness and enables long range magnetic order.} The 
multi-band character of the semiconductor valence plays an essential role 
in the ferromagnetism of these materials.

\subsection{Limits on the Curie temperature}

Isotropic ferromagnets have spin-wave Goldstone collective modes whose 
energies vanish at long wavelengths,
\begin{equation}
\label{long-wavelength}
   \Omega_{k} = D k^2 + {\cal O} (k^4)\, ,
\end{equation}
where $k$ is the wavevector of the mode.
Spin-orbit coupling breaks rotational symmetry which leads to a finite gap,
see Eq.~(\ref{gap+stiffness}).
According to our numerical studies, though, this gap is negligibly small as far
as the suppression of ferromagnetism by collective spin excitations is 
concerned and can, therefore, be dropped for the present discussion.
Each spin-wave excitation reduces the total spin of the ferromagnetic state 
by 1. 
The coefficient $D = 2A / (N_{\rm Mn}S)$ is proportional to the spin stiffness
$A$.
These collective excitations are not accounted for in the mean-field 
approximation.   
If the spin stiffness is small, they will dominate the suppression of the 
magnetization at all finite temperatures and limit the critical temperature. 
In this case, the typical local valence-band carrier polarization remains
finite above the critical temperature.
Ferromagnetism disappears only because of the loss of long-range spatial 
coherence. 

A rough upper bound on the critical temperature $T_c^{\rm coll}$ can
be obtained by the following argument which accounts for the role of 
collective fluctuations \cite{schliemannapl01}.  
The magnetization vanishes at the temperature where the number of excited spin 
waves equals the total spin of the ground state.  
\begin{equation}
   N_{\rm Mn}S = {1\over 2 \pi^2} \int_0^{k_D} dk\, k^2 n(\Omega_{k}) \, ,
\end{equation}
where $n(\Omega_{k})$ is the Bose occupation number and the Debye cutoff,
$k_D=(6\pi^2N_{\rm Mn})^{1/3}$, ensures the correct number of magnetic ion
degrees of freedom.
We therefore find that the critical temperature of a ferromagnet cannot exceed 
\begin{equation}
   k_BT_c^{\rm coll} = {2S+1\over 6} D k_D^2
\label{collective_tc}
\end{equation}
for $S\ge 5/2$ where $D$ is the $T=0$ spin-stiffness.
To obtain this equation, we have assumed that the spin waves can be
approximated as non-interacting Bose particles, replaced the dispersion by the 
long-wavelength limit Eq.~(\ref{long-wavelength}), and noted that the critical 
temperature estimate is proportional to $D k_D^2$, justifying the use of the 
classical expression for the mode occupation number 
$n_{k} \approx k_B T/\Omega_k - 1/2$. 
These considerations set an upper bound on the critical temperature which is 
proportional to the spin stiffness, a bound not respected by mean-field theory.

To get a qualitative but transparent picture we employ the two-band model with
parabolic bands, and deduce the spin stiffness from Eqs.~(\ref{eq:sw_strong})
and (\ref{eq:sw_weak}) for the strong and weak-coupling regime, respectively.
For strong coupling, $\Delta/\epsilon_F \gg 1$, the exchange coupling 
completely polarizes the valence-band electrons, and we find (using 
$p \ll 2N_{\rm Mn}S$) the $T_c$ bound
\begin{equation}
\label{coll_s}
   T_c^{\rm coll,s} = {2S+1\over 12S}\epsilon_F 
        \left({p\over N_{\rm Mn}}\right)^{1/3} \, . 
\end{equation}
For small $\Delta/\epsilon_F$, the weak-coupling or RKKY regime, exchange 
coupling is a weak perturbation on the band system. 
In this regime we get
\begin{equation}
\label{coll_RKKY}
   T_c^{\rm coll,RKKY} = T_c^{\rm MF} {2S+1 \over 12(S+1)\sqrt[3]{2}}
        \left({N_{\rm Mn}\over p}\right)^{2/3} \, ,
\end{equation}
i.e., mean-field theory is reliable only for $p/N_{\rm Mn} \ll 1$, as expected 
since in this case the RKKY interaction has a range which is long compared to 
the distance between Mn spins. 
\begin{figure}
\centerline{\includegraphics[width=8cm]{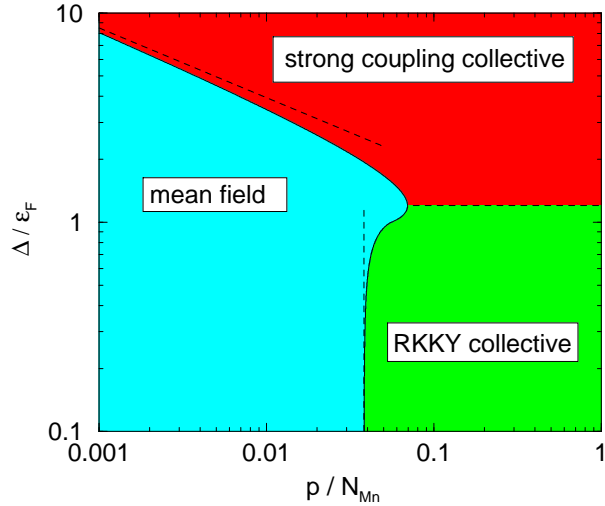}}
\caption{Critical-temperature-limit regimes for the two-band model.  
        In the mean-field regime $T_c$ is limited by individual Mn spin 
        fluctuations. 
        In the collective regimes, the critical temperature is limited by 
        long-wavelength fluctuations with a stiffness proportional to the 
        bandwidth for weak (RKKY) exchange coupling and inversely proportional 
        to the bandwidth for strong exchange coupling. 
        At the solid line $T_c^{\rm MF}=T_c^{\rm coll}$. 
        Dashed lines: expansions for large and small $\Delta/\epsilon_F$, 
        Eqs.~(\ref{coll_s}) and (\ref{coll_RKKY}), and the crossover from the 
        RKKY to the strong coupling collective regime. }
\label{phases}
\end{figure}

We expect that the qualitative picture derived from the two-band model will 
persist for the six-band model.
The actual values of the boundaries between the regimes indicated in 
Fig.~\ref{phases} will, however, be shifted because of the important 
differences in the microscopic physics that determines the spin stiffness
of the two models discussed above.  We expect that the $T_c$ estimates 
derived in the preceding paragraph will be directly applicable to n-type
carrier-mediated ferromagnets.   {\em As a consequence we expect that it
will be impossible to achieve large ferromagnetic transition temperatures 
in n-type semiconductors with carrier mediated ferromagnetism.}
The $\sim5\%$ reduction of (Ga,Mn)As mean-field $T_c$ due to spin fluctuations,
mentioned in Section~4.1, was obtained using the six-band model and solving 
self-consistently Eq.~(\ref{collective_tc}) and 
$D(T_{c}^{\rm coll})=D(T=0)\langle S\rangle(T_{c}^{\rm coll})/S$.
Larger reductions compared to mean-field-theory estimates are expected in
some hosts, but mean-field-theory retains a qualitative validity.

%% file: john.tex
\section{Collective fluctuations beyond spin wave theory
and continuum approximation}
\label{section_mc}

In the preceding sections we described the magnetic properties of
Mn-doped semiconductors by a model of itinerant carriers
which are exchange-coupled to localized magnetic moments formed by the
dopants. An important property, resulting from the growth process of such materials,
is that the Mn acceptors are distributed at random on the cation sites of the
underlying crystal lattice.  To this point in this chapter, we have used a continuum
approximation for the Mn ion distribution that yields a disorder-free problem.
The continuum approximation makes it possible to obtain some results  
analytically, and also crucially simplifies numerical calculations.
However, it neglects substitutional disorder in the Mn positions which
can have a substantial impact on the
ferromagnetism. That this is true is evident from the fact that
magnetic properties of presently available samples are often  
sensitive to the conditions of their fabrication, and reproducibility is
achieved only if the growth parameters are carefully controlled
\cite{ohnojmmm99}. Moreover, recent studies of post-growth annealed
(Ga,Mn)As samples have revealed that the magnetic \cite{Schiffer} as 
well as the structural \cite{Molenkamp} properties can depend crucially 
on the type of defects and disorder present in the system.

The spin-wave theory presented in the previous section
describes collective excitations of the ion spin system (modeled as a 
continuum) in terms of Gaussian fluctuations around the ferromagnetic
ground state in the many-body path integral. This
non-interacting spin-wave theory is exact at low temperatures, where 
deviations from the ordered ground state are small, but is less
reliable when the temperature is raised
toward the ferromagnetic transition temperature. 

In this section we complement the theoretical approaches described above by
methods which (i) take disorder effects due to the
randomly chosen ion positions into account and (ii) are able to address directly the
region of larger deviations of the spin configuration from the
ferromagnetically ordered state. Specifically, we present results of
Monte Carlo simulations \cite{schliemannapl01,schliemanncm33}
which treat the minimal model without approximation.
Finally we report on a rigorous stability analysis of the
perfectly ferromagnetically ordered {\em collinear state} of Mn ions in the
presence of disorder\cite{schliemanncm0107573}.
We predict that {\em noncollinear ferromagnetism is common in (III,Mn)V semiconductors};
the robust collinear ferromagnetic states that have the highest ferromagnetic 
transition temperatures occur only when the carriers are relatively weakly
localized around individual Mn ions.

\subsection{Model considerations}

The considerations in this section are based on the kinetic-exchange model 
discussed in the previous
sections. To account for the finite extent of the Mn ions 
\cite{Bhatta:00}
in the exchange term we replace
$J_{\rm pd} {\vec s}({\vec R}_I) \cdot {\vec S}_I$ by 
$\int d^3 r J ( {\vec r} - {\vec R}_I) {\vec s}({\vec r})\cdot 
{\vec S}_I$ with the finite-range exchange parameter
\begin{equation}
  J(\vec r )=\frac{J_{\rm pd}}{(2\pi a_{0}^{2})^{\frac{3}{2}}}
  e^{- r^{2}/(2 a_{0}^{2})}\,.
\label{excpl}
\end{equation}
Both the strength $J_{\rm pd}$ and range $a_0$ of this interaction are 
phenomenological parameters to be fixed by comparison with experiment or, 
ideally, to be extracted from first principles electronic structure 
calculations.
Note that the exchange-coupling range parameter $a_0$ in Eq.~(\ref{excpl}) is required 
in our calculations once the discreteness of the Mn ions is acknowledged; exchange-coupling shifts
of quasiparticle energies would diverge otherwise.
Given this finite range, the only approximation we make below in treating
the minimal model is that we treat the Mn spins classically.  Because of the 
relatively large value of the Mn ion spins, this approximation should have 
minimal consequence except for the leading low temperature magnetization suppression.

We are interested in thermal expectation values of the form 
\begin{equation}
\bar f =
\frac{1}{\cal Z}
\int_{0}^{2\pi}d\varphi\int_{0}^{\pi}d\vartheta
\sin\vartheta\,
{\rm Tr}\left\{\hat f(\vartheta,\varphi)e^{-\beta{\cal H}}\right\}\,,
\label{exval1}
\end{equation}
where $\beta$ is the inverse temperature, $\cal Z$ the partition function,
and $\vartheta$, $\varphi$ are
shorthand notations for the whole set of classical spin coordinates.
The quantity $\hat f(\vartheta,\varphi)$ is a function of the ion spin
angles and an operator with respect to the quantum mechanical carrier degrees
of freedom over which the trace is performed.
In practice we replace the fermion trace by a ground state 
expectation value, since the temperatures of interest will always be much
smaller than the Fermi energy.  For typical carrier 
densities $p$ of order 0.1~nm$^{-3}$,
the Fermi temperature for the carriers is typically larger than $1000{\rm K}$,
compared to ferromagnetic critical temperatures $ \sim 100{\rm K}$.
Thermal effects in the carrier system are therefore negligible.  Thus, 
\begin{equation}
\bar f = \frac{1}{\cal Z}
\int_{0}^{2\pi}d\varphi\int_{0}^{\pi}d\vartheta
\sin\vartheta\,
\langle 0|\hat f(\vartheta,\varphi)|0\rangle
e^{-\beta\langle 0|{\cal H}|0\rangle}\,,
\label{exval2}
\end{equation}
where $|0\rangle$ denotes the groundstate of non-interacting 
fermions with the appropriate band Hamiltonian and a Zeeman-coupling term 
$h$ whose effective magnetic field $H_{\rm eff}$ is due to exchange 
interactions with the localized spins,
\begin{eqnarray}
h &=& g \mu_B \int d^{3}r\, \vec s(\vec r ) \cdot \vec H_{\rm eff}(\vec r ) 
\nonumber \\
\vec H_{\rm eff}(\vec r ) 
&=& \sum_{I} J(\vec r - {\vec R}_{I}) S  {\hat \Omega}_I / (g \mu_B)
\label{eq:beff}
\end{eqnarray}
where $\hat\Omega_I 
=(\sin \theta_I \cos \phi_I ,\sin \theta_I \sin \phi_I ,\cos \theta_I)$
is the direction of the classical spin at ${\vec R}_I$.
In the following we denote thermal expectation values of quantities defined 
in terms of classical spin orientation variables 
by $\langle\cdot\rangle$ and quantum mechanical expectation values within 
the carriers ground state by $\langle 0|\cdot|0\rangle$. 

\subsection{Remarks on the Monte Carlo method}
\label{HMCalg}

A standard way to evaluate expectation values of the form Eq.~(\ref{exval2}) 
is to use classical Monte Carlo algorithms which perform a random walk in 
phase space of the classical variables ($\vartheta$,$\varphi$). 
The probabilities governing this
Monte Carlo dynamics are specified by the dependence of many-fermion
energy on the localized-spin configuration.
The many-fermion ground state is a Slater determinant whose single-particle 
orbitals are the lowest energy eigenstates of a single-band or multi-band 
Hamiltonian. For the case of a parabolic
band, the matrix elements of the corresponding one-particle Hamiltonian 
in a plane-wave basis read
\begin{equation}
\langle\vec k^{\prime} \sigma^{\prime}|{\cal H}|\vec k \sigma\rangle 
=\frac{\hbar^{2}k^{2}}{2m^{\ast}}\delta_{\vec k^{\prime} \vec k}
\delta_{\sigma^{\prime} \sigma} 
+\frac{S}{2L^{3}}\sum_{I}J_{\vec k-\vec k^{\prime}}
e^{i(\vec k-\vec k^{\prime})\vec R_{I}}\hat\Omega_{I} \cdot
\vec\tau_{\sigma' \sigma}\,,
\label{singpart}
\end{equation}
where $\vec k$ and $\sigma$ denote wavevector and spin indices, respectively,
$J_{\vec k}$ is the Fourier transform of $J(\vec r)$, and $L$ the edge 
length of the simulation cube. Periodic boundary conditions restrict the 
admissible values of wavevector components to integer multiples 
of $2\pi/L$. In Eq.~(\ref{singpart}), $\vec\tau$ is the vector of Pauli
spin matrices. 

Since the many-particle ground state of the carrier system has to be
computed at each Monte Carlo step, the computational effort required for
the present calculations is much larger 
than in simple classical spin models.  In the usual Metropolis algorithm,
a single spin orientation is altered at each step.  If this algorithm
were employed here, the time required to diagonalize the single-particle
Hamiltonian each time would severely limit the efficiency of the algorithm. 
We therefore use the Hybrid Monte Carlo algorithm, which was introduced in 
the mid 1980's in the context of lattice field theories \cite{HMC}.
In this method {\em all} classical variables are altered in one Monte Carlo 
step.
This drastically reduces the number of matrix diagonalizations required to 
explore statistically important magnetic configurations.
The Hybrid algorithm is a powerful method for Monte Carlo simulations in 
systems containing coupled classical and quantum mechanical degrees of freedom.

A concrete Monte Carlo simulation necessarily works in a system of finite
size.  In the present case, calculation of the fermion ground state 
also requires truncation of the plane-wave expansion we use for the independent 
particle wavefunctions that diagonalize the Hamiltonian (\ref{singpart}).
We are able to obtain convergence with respect to this
truncation only when the range of the exchange interaction $a_0$ is not too short.
taking into account a finite number of plane-wave states entering the
single-particle carrier Hamiltonian (\ref{singpart}). The importance of these
finite-size effects in real and reciprocal space (using periodic boundary conditions),
and their interplay with the regularization parameter $a_{0}$, are discussed in detail in
Ref.~\cite{schliemanncm33}. In the following we shall leave aside such 
technical aspects and concentrate on the physical results.

\subsection{Numerical Monte Carlo Results}
\label{numres}

In this subsection we present numerical Monte Carlo results.  
We concentrate on the spin 
polarizations of the Mn ions and the carriers as a function of 
temperature and address the ferromagnetic transition. 
We start with the two-band model. 

\subsubsection{Two-band model.}

\begin{figure}
\centerline{\includegraphics[width=8cm]{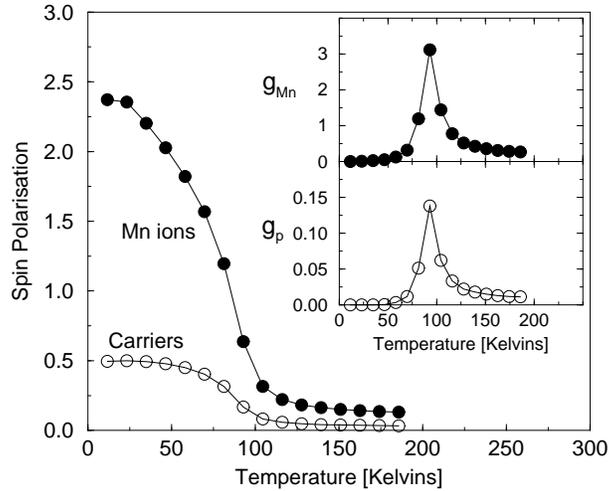}}
\caption{Magnetization curves for Mn ions and carriers.
The upper and lower inset show the magnetic fluctuations for the Mn
ions and the carriers, respectively. Both differ by a factor of approximately
25 reflecting the square of the ratio of spin lengths.
The density of Mn
ions is $N_{\rm Mn}=1.0{\rm nm^{-3}}$, the carrier density is 
$p=0.1{\rm nm^{-3}}$ in a cubic volume of $V=540{\rm nm^{3}}$.
The band mass is half the bare electron mass with an exchange parameter of
$J_{\rm pd}=0.15{\rm eVnm^{3}}$.
\label{fig_mc_1}}
\end{figure}

Fig.~\ref{fig_mc_1} shows typical magnetization data as a function of
temperature. These results were obtained for a Mn ion density of
$N_{\rm Mn}=1.0{\rm nm^{-3}}$, a carrier density $p=0.1{\rm nm^{-3}}$ in a cubic 
simulation volume of $V=540{\rm nm^{3}}$, i.e., the system contains 
540 Mn
ions and 54 carriers. The effective band mass is half the bare electron mass,
and the chosen exchange parameter is $J_{\rm pd}=0.15{\rm eVnm^{3}}$.
Here and in the following Monte Carlo data the range parameter for the
exchange coupling is $a_{0}=0.1{\rm nm}$.
 
The main panel shows the average polarization of the Mn spins,
\begin{equation}
M=\frac{1}{N_{\rm Mn}V}\langle|\vec S_{\rm tot}|\rangle\,,
\end{equation}
i.~e. the thermally averaged modulus of the total ion spin, along with
the carrier magnetization,
\begin{equation}
m=\frac{1}{pV}\langle|\langle 0|\vec s_{\rm tot}|0\rangle|\rangle\,,
\label{carrmag}
\end{equation}
which is the ensemble average of the modulus of the total ground-state carrier 
spin. Both quantities are divided by the number of 
particles and are close to their maximum values at low temperatures.
At higher temperatures they show the expected transition to a paramagnetic 
phase. The critical temperature of this ferromagnetic transition 
is most readily estimated from numerical results for the magnetization
fluctuations:
\begin{eqnarray}
g_{\rm Mn} & = & \frac{1}{N_{\rm Mn}V}\left(\langle|\vec S_{\rm tot}|^{2}\rangle
-\langle|\vec S_{\rm tot}|\rangle^{2}\right)\,,
\label{fluc1}\\
g_{p} & = & \frac{1}{pV}\left(
\langle|\langle 0|\vec s_{\rm tot}|0\rangle|^{2}\rangle
-\langle|\langle 0|\vec s_{\rm tot}|0\rangle|\rangle^{2}\right)\,.
\label{fluc2}
\end{eqnarray}
These two fluctuations per particle are plotted in the insets of 
Fig.~\ref{fig_mc_1}. 
They both show a pronounced peak at a temperature $T \sim 100 {\rm K}$,
defining the finite-system transition temperature for these model 
parameter values.  
In fact, in a region around this transition the two data sets differ by 
a factor of approximately 25, which is the square of the ratio of the two 
spin lengths entering the expressions (\ref{fluc1}) and (\ref{fluc2}), 
respectively.  This observation shows explicitly that
the correlation length is the same for Mn ions and the carrier 
system near the transition, that is both approach the finite 
system size of the simulation.

Our Monte Carlo approach clearly reproduces the expected 
ferromagnetic transition.  {\em Ferromagnetism still occurs
when randomness in the Mn positions is accounted for.}  
The transition temperature $T_c$ can be determined unambiguously and consistently 
from the positions of very pronounced peaks in total magnetization 
fluctuations of both the Mn ions and the carriers.

\subsubsection{Results for $T_{c}$.}

We now turn to the transition temperature $T_{c}$ for the two-band
model. Within mean-field theory this quantity is given by Eq.~(\ref{tc}).
Our objective here is not to make a quantitative prediction of
the critical temperature for particular ferromagnetic semiconductor 
systems.  By doing a numerically exact calculation for 
a model that captures much of the physics, however, we hope to
shed light on the range of validity and the sense and 
magnitude of likely corrections to mean-field-theory $T_c$ estimates.

The mean-field expression for $T_{c}$ can be obtained by averaging the ion-spin 
and carrier polarizations over space.  The effective field which each  Mn
spin experiences due to a finite carrier polarization is then constant in space and
the carrier bands are in turn rigidly spin split by $\Delta=J_{\rm pd}N_{\rm Mn}S$. 
(The limit in which mean-field theory is exact can be achieved in our model 
by letting $a_0 \rightarrow \infty$ in Eq.~(\ref{excpl}).)
Mean-field theory, which neglects spatial fluctuations and correlations 
between carriers and Mn spins, predicts that $T_{c}$ is  
quadratic in the exchange parameter $J_{\rm pd}$ and linear in the effective 
band mass $m^*$, when correlations in the itinerant system are neglected.

Here we determine the critical temperature $T_{c}$ with the help of
our Monte Carlo scheme. 
\begin{figure}
\centerline{\includegraphics[width=8cm]{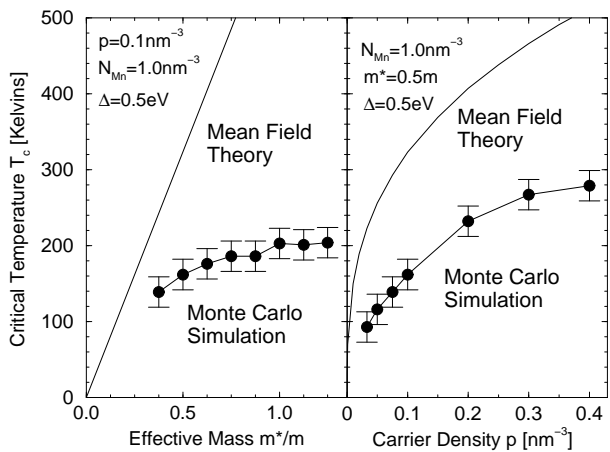}}
\caption{The critical temperature $T_{c}$ as a function of the carrier mass
(left panel) and the carrier density (right panel).
The exchange parameter is in both cases $J_{\rm pd}=0.2{\rm eVnm^{3}}$
leading to a zero temperature mean-field spin splitting
of $\Delta=J_{\rm pd}N_{\rm Mn}S=0.5{\rm eV}$.
The results of the Monte Carlo runs are compared with the mean-field
predictions.
\label{fig_mc_2}}
\end{figure}
In Fig.~\ref{fig_mc_2} we show results for  Mn densities 
$N_{\rm Mn}=1.0{\rm nm^{-3}}$ and a mean-field band splitting 
$\Delta=J_{\rm pd}N_{\rm Mn}S=0.5{\rm eV}$. 
The left panel shows the dependence of the critical temperature on the 
carrier effective mass. 
This dependence is very important for the search for diluted magnetic 
semiconductor systems with $T_{c}$'s larger than room temperature. 
In the mean-field approximation, $T_{c}$ grows linear with increasing mass. 
The Monte Carlo results clearly deviate from this prediction suggesting a 
saturation of $T_{c}$ at carrier masses close to the bare electron mass. 
For even higher masses, we expect the electrons to behave more classically 
and localize around individual Mn spins, suppressing long-range order further.   
In this limit the electronic energy cost of changes in the relative orientations of nearby
Mn spins will get smaller causing $T_c$ to decline, and eventually 
ferromagnetism will disappear.  
This is consistent with the observation that, within the continuum model,
the spin stiffness declines as $1/m^*$ for large band masses
\cite{koenigprl00,koenigspringer01,schliemannapl01}.  As mentioned earlier, 
the spin stiffness tends to be remain substantially larger when coupled
light and heavy holes are retained in the calculation.  Still, this 
calculation demonstrates that mean-field-theory must be regarded 
with some caution and its validity must be checked in each new circumstance.

To discuss the critical temperature as a function of the exchange coupling 
parameter $J_{\rm pd}$, we observe that the Hamiltonian of itinerant carriers
satisfies the scaling relation
\begin{equation}
\beta{\cal H}\left(m^{\ast},J_{\rm pd}\right)
=\frac{\beta}{q}{\cal H}\left(\frac{m^{\ast}}{q},q J_{\rm pd}\right)
\label{scalerel}
\end{equation}
with $q>0$.
Therefore the saturation of $T_{c}$ as a function of the effective mass at 
fixed $J_{\rm pd}$ corresponds to a linear dependence of $T_c$ on $J_{\rm pd}$ 
at fixed $m^{\ast}$. This contrasts with the mean-field prediction
$T_c^{MF} \propto J_{\rm pd}^2$.

In the right panel of Fig.~\ref{fig_mc_2} we show $T_{c}$ as a function of the 
carrier density. 
Here the Monte Carlo approach also clearly yields a lower critical
temperatures than mean-field theory. For still higher carrier densities
the typical distance between nearby Mn ions will
become larger than the band electron Fermi wavelength, causing the sign
of the typical exchange coupling to oscillate in an RKKY fashion.
As we shall see in the next subsection, the
resulting frustration makes the ferromagnetic state
unstable, possibly leading to a regime of spin-glass order.  
Disordered states can also occur when the exchange coupling becomes strong.

\subsubsection{Six-band model.}

\begin{figure}
\centerline{\includegraphics[width=8cm]{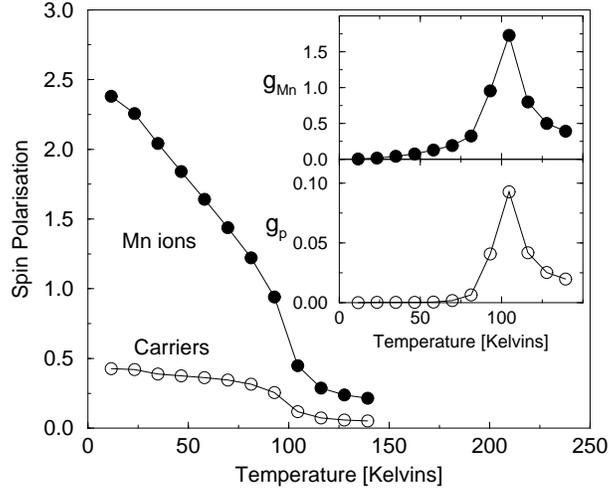}}
\caption{Magnetization curves for Mn ions and carriers in the six-band model
for n exchange coupling of $J_{\rm pd}=0.15{\rm eVnm}^{3}$. The carrier density is
$p=0.1{\rm nm}^{-3}$ with an Mn ion density of $N_{\rm Mn}=1.0{\rm nm}^{-3}$
in a volume of $V=280{\rm nm}^{3}$. As in the case of parabolic bands,
the ferromagnetic transition in clearly and consistently signaled
by pronounced peaks in the magnetic fluctuations shown in the insets.
\label{fig_mc_3}}
\end{figure}

We now turn to the six-band model in which the kinetic energy part is given by 
the Kohn-Luttinger model.
Fig.~\ref{fig_mc_3} shows typical magnetization data for the (Ga,Mn)As 
system, assuming 
exchange coupling $J_{\rm pd}=0.15{\rm eVnm}^{3}$, carrier density is
$p=0.1{\rm nm}^{-3}$, and Mn ion density $N_{\rm Mn}=1.0{\rm nm}^{-3}$
in a volume of $V=280{\rm nm}^{3}$. 
As in the case of parabolic bands, a 
ferromagnetic transition is clearly signaled by pronounced peaks 
in the magnetic fluctuations of both Mn ions and carriers.
We find that, in contrast to the parabolic two-band model, the carrier 
magnetization is already reduced at temperatures well below $T_{c}$.
In fact, because of strong spin-orbit coupling in the valence band, 
full polarization of the carrier spins never occurs.
Another difference compared to the parabolic-band model concerns the shape of 
the magnetic fluctuations for the Mn ions and the carriers
as a function of temperature.
Although both curves indicate the same value for $T_{c}$ for these parameter values
, their shape in the vicinity of $T_c$ is slightly different and the ratio of 
their fluctuations is smaller than $25$ (the square of the ratio of
spin lengths involved).  These differences arise because of the more complicated band structure
and the spin-orbit coupling present in the Kohn-Luttinger Hamiltonian.
\begin{figure}
\centerline{\includegraphics[width=8cm]{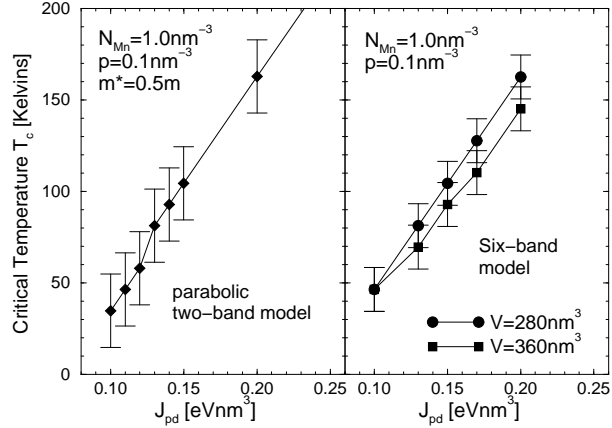}}
\caption{The right panel shows critical temperature $T_{c}$ as a function of 
the exchange parameter $J_{\rm pd}$ for the same particle densities as in
Fig.~\ref{fig_mc_3} and two different system sizes. Both data sets agree within
error bars and show a linear dependence of $T_{c}$ on $J_{\rm pd}$.
In the left panel, the corresponding data for the parabolic two-band model
with an effective mass of half the bare electron mass is plotted.
The latter system is a reasonable approximation to the six-band case.
\label{fig_mc_4}}
\end{figure}   
In the right panel of Fig.~\ref{fig_mc_4} we plot the transition temperature 
as a function of the exchange coupling $J_{\rm pd}$ for two different system 
sizes. Both data 
sets agree within error bars and show a linear dependence of $T_{c}$ on 
$J_{\rm pd}$.  This finding is the same as for the two-band model and contrasts 
with mean-field theory which predicts $T_c \propto J_{\rm pd}^2$. The left panel
of Fig.~\ref{fig_mc_4} shows the transition temperature as a 
function of $J_{\rm pd}$ for the parabolic model for an effective mass
$m^{\ast}=0.5m_e$. This value is close to the heavy-hole mass in the
Kohn-Luttinger model for parameters appropriate for GaAs.
The data in the left panel can be obtained from the left panel of 
Fig.~\ref{fig_mc_2} via the scaling relation (\ref{scalerel}).
Comparing the two panels of Fig.~\ref{fig_mc_4} demonstrates that, in the 
range of carrier densities studied here, a single parabolic band with an
effective mass close to that of the heavy-hole Kohn-Luttinger-model band 
provides a reasonably good approximation to the behavior of the
six-band system.  We expect larger differences to occur in the strong coupling 
regime, which presents technical difficulties to the Monte Carlo calculations.

\subsection{Disorder effects and noncollinear ferromagnetism}

\begin{figure}
\centerline{\includegraphics[width=8cm]{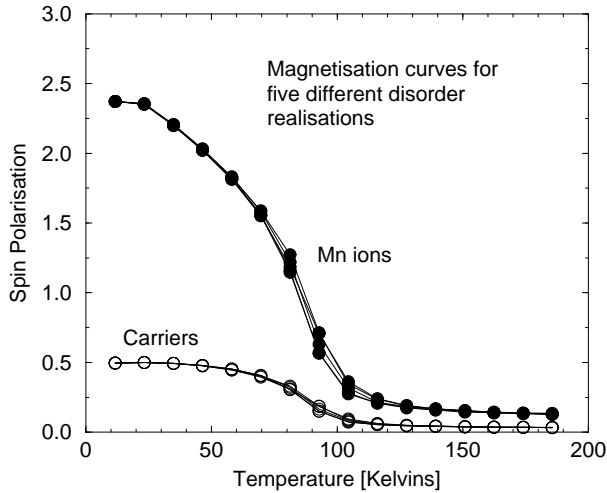}}
\caption{Magnetization curves for five different realizations of Mn
positions underlying the averaged data of Fig.~\protect{\ref{fig_mc_1}}.
\label{fig_mc_5}}
\end{figure}

The magnetization data of
Fig.~\ref{fig_mc_1} were obtained by averaging over five different
realizations of the Mn ions. In fact over the range of parameters we 
have explored, except for the larger values of 
$p/N_{\rm Mn}$, results for 
different disorder realizations differ only very weakly from each other.
This is illustrated in Fig.~\ref{fig_mc_5}, where the magnetization curves
underlying the averaged results of Fig.~\ref{fig_mc_1} are plotted.
Those five datasets are hardly distinguishable from each other.
Positional disorder in the localized magnetic moments of Mn
ions can however affect the nature of the ground state itself in some
circumstances. As seen
from Fig.~\ref{fig_mc_1}, the ion spin magnetization seems to saturate
at low temperatures at a value slightly smaller than its maximum, given
by the Mn spin length of $5/2$. Moreover, as shown in 
Fig.~\ref{fig_mc_5}, this behavior occurs quite consistently for 
different disorder realizations.  As we see below, the effect is
due to the randomness in Mn positions which,
combined with the RKKY-like oscillations in the effective coupling between
Mn spins, makes the perfectly ferromagnetically ordered {\em collinear state}
state unstable and leads to {\em noncollinear ferromagnetism}.

We now outline a theory of magnetic fluctuations around a given state
within a path integral formalism similar to the spin-wave approach
presented in Section \ref{section_sw}.2 for fluctuations around the 
collinear ferromagnetic state. In contrast to the spin-wave calculations,
here we do not
approximate the Mn magnetic moments by a continuum but retain them,
as in the Monte Carlo approach, as 
individual spins of length $S=5/2$ placed at arbitrary locations.
Let us first consider magnetic fluctuations around the perfectly
ferromagnetically ordered collinear state having all Mn spins 
oriented in parallel.

Repeating the steps described in Section \ref{section_sw}, but keeping the
Mn spins as individual objects, one obtains the following expression
for the fluctuation part of the effective action in up to second order in the 
bosonic spin variables:
\begin{equation}
{\cal S}_{\rm eff}=\frac{1}{\beta}\sum_{m}\sum_{I,J}
\bar z_{I}(\nu_{m})D^{-1}_{IJ}(\nu_{m})z_{J}(\nu_{m})
\label{flucaction}
\end{equation}
where $\nu_{m}=2\pi m/\beta$ is a Matsubara frequency. 
$z_{I}(\nu_{m})$ stands for the bosonic Holstein-Primakoff field of 
Mn spin $I$ that  describes deviations from a fully aligned state. 
The fluctuation matrix $D^{-1}_{IJ}(\nu_{m})$ reads
\begin{equation}
D^{-1}_{IJ}(\nu_{m})=L_{IJ}(\nu_{m})+K_{IJ}(\nu_{m})
\end{equation}
with
\begin{eqnarray}
L_{IJ} & = &\delta_{IJ}\Big(-i\nu_{m}-
 \int d^{3}rJ(\vec r-\vec R_{I})\langle s^{z}(\vec r)\rangle\Big)\; ,\\
K_{IJ} & = & \frac{S}{2}\sum_{\alpha,\beta}\Biggl[
\frac{f(\eta_{\alpha})-f(\eta_{\beta})}
{i\nu_{m}+\eta_{\alpha}-\eta_{\beta}}
F^{\alpha\downarrow,\beta\uparrow}_{I}
F^{\beta\uparrow,\alpha\downarrow}_{J}\Biggr]\; .
\label{K}
\end{eqnarray}
Here $\langle\vec s(\vec r)\rangle$ is the expectation value of the
carrier spin density, $f$ the Fermi function, and 
\begin{equation}
F^{\alpha\sigma,\beta\mu}_{I}=\int d^{3}rJ(\vec r-\vec R_{I})
\bar\psi_{\alpha\sigma}(\vec r)\psi_{\beta\mu}(\vec r)
\label{F}
\end{equation}
with $\psi_{\alpha\sigma}(\vec r)$ being the spin component $\sigma$ of the 
carrier wave function with label $\alpha$ and energy $\varepsilon_{\alpha}=
\eta_{\alpha}+\mu$, where $\mu$ is the chemical potential for the carriers
All quantities referring to the carrier system are to be
evaluated for the collinear orientation of Mn spins.

The zero-frequency ($m=0$) contribution to the effective action
(\ref{flucaction}) describes the energy of static fluctuations around
the collinear state.  For this state to be stable, the matrix
$D^{-1}_{IJ}(0)$ must have non-negative eigenvalues only, while the
occurrence of negative eigenvalues of this matrix indicates that the
perfectly collinear state is not the ground state. This interpretation of
the zero-frequency fluctuation term is confirmed by the observation that
this contribution is also obtained (at zero temperature) 
by a standard perturbation
theory for the carrier ground state energy with respect to small
deviations of the ion spins from collinear orientation. The formalism
given above embeds this finding in a more general theory of dynamic
fluctuations at finite temperature. However, we shall
concentrate in the following on static ground state properties, i.e. on $T=0$, where
the Fermi functions become step functions. 

We note that for any
arrangement of the Mn positions $R_{I}$, the matrix $D^{-1}_{IJ}(0)$ contains
a zero eigenvalue corresponding to a uniform rotation of all spins.
The eigenvalues of $D^{-1}_{IJ}(0)$ are proportional to
magnetic excitation energies. In this sense the eigenvalue
distribution of $D^{-1}_{IJ}(0)$ can be
interpreted as a density of states  for magnetic excitations.

We have evaluated the spectrum of $D^{-1}_{IJ}(0)$ in systems given by
a simulation cube with periodic boundary conditions averaging over different
realizations of the Mn positions. The single-particle wavefunctions
$\psi_{\alpha\sigma}(\vec r)$ are computed in a plane-wave basis taking into
account wavevectors $\vec q$ with length up to an appropriate cutoff $q_{c}$.
The same truncated plane-wave basis is used to compute the quantities
(\ref{F}) entering (\ref{K}). Note that, for fluctuations around the
collinear ferromagnetic state, $D^{-1}_{IJ}(i\omega)$ is always real
and symmetric for real $\omega$ since all carrier wavefunctions 
have, for a given spin projection $\sigma$, the same coordinate-independent
phase. This follows from the fact
that the single-particle Hamiltonian describes for each spin projection 
just the problem of a spinless particle in a potential landscape provided by 
the Mn ions. Since $D^{-1}_{IJ}(i\omega)$ is real and symmetric, the
components of each of its eigenvectors all have the same phase
(and can be chosen to be real). Physically this corresponds to the
invariance of the system under rotations around the magnetization axis of
the collinear state.

\begin{figure}
\centerline{\includegraphics[width=8cm]{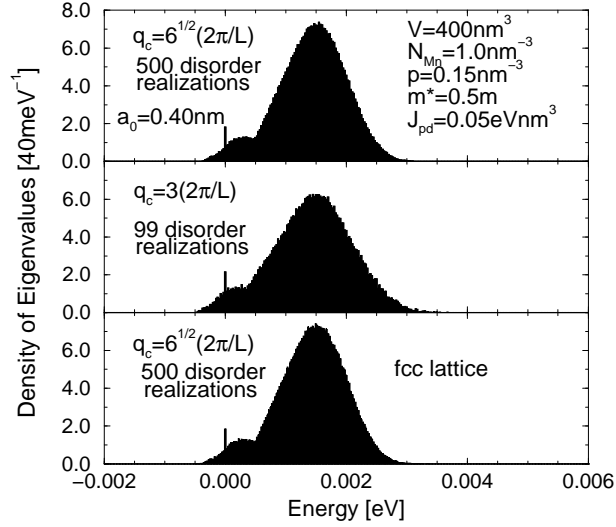}}
\caption{The disorder-averaged eigenvalue density of the matrix
$D^{-1}_{IJ}(0)$ describing magnetic fluctuations around the collinear
state. The data are obtained  
for a simulation cube of volume $V=L^{3}=400{\rm nm}^{3}$ with a Mn density 
of $N_{\rm Mn}=1.0{\rm nm}^{-3}$ 
and a density of $p=0.15{\rm nm}^{-3}$ of carriers having a band mass
of half the bare electron mass. The strength of the exchange interaction
between ions and carriers is $J_{pd}=0.05{\rm eVnm}^{-3}$ with a spatial
range of $a_{0}=0.40{\rm nm}$. The two upper panels show data for different
wavevector cutoff $q_{c}$ with the Mn positions chosen completely at 
random. The lowest panel contains data for the same situation as the
top one but with the Mn positions chosen from an fcc lattice.
The peaks at zero energy are due to the uniform 
rotation mode which strictly occurs in any disorder realization.
\label{fig_mc_6}}
\end{figure}

The two upper panels of
Fig.~\ref{fig_mc_6} show results for typical system parameters for two
different values of $q_{c}$.  Comparison of the panels shows that the
effects of the wavevector cutoff on the low-lying excitations have already
saturated for the smaller $q_{c}$.  Almost all eigenvalues of $D^{-1}_{IJ}(0)$ 
lie at positive energies, but there is often a small fraction of negative eigenvalues,
indicating an instability of the collinear state.

In the calculations discussed so far, the Mn positions were chosen 
completely at random with a uniform distribution, while in a real
(III,Mn)V semiconductor the Mn ions are supposed to be located on
the cation sites forming an fcc lattice. In the bottom panel of 
Fig.~\ref{fig_mc_6} we show data for the same system parameters as in the
top panel but with the Mn positions chosen from an appropriate fcc
lattice such that about 5\% of all sites are occupied. Both plots
are practically identical indicating that our observations do not depend
on this detail of the modeling.

The shape of the eigenvalue distribution of the fluctuation matrix
$D^{-1}_{IJ}(0)$ is, in the model we have studied,
sensitive  to the Mn density $N_{Mn}$, the carrier density $p$,
and the Hamiltonian parameters $m^{\ast}$, $J_{pd}$, and
$a_{0}$. Situations in which the collinear ferromagnetic state
is, for certain disorder realization, stable can be approached
most simply, for technical reasons, by letting $a_0$ be larger.
However for the value of $p/N_{Mn}$ illustrated in Fig.\ref{fig_mc_6},
corresponding to a carrier density somewhat lower than measured in
the highest $T_c$ samples, negative eigenvalues occur for nearly any Mn
ion distribution.
We note that negative eigenvalues increase in number as the 
wavevector cutoff is increased toward its converged value.  
These results suggest that noncollinear ferromagnetic states are common, 
that they are sensitive to 
the distribution of Mn ions and other defects - 
especially those that trap carriers, and that the collinear ferromagnetic 
state tends to become unstable as mean-field band eigenfunctions become 
more strongly localized around Mn ion site

\begin{figure}
\centerline{\includegraphics[width=8cm]{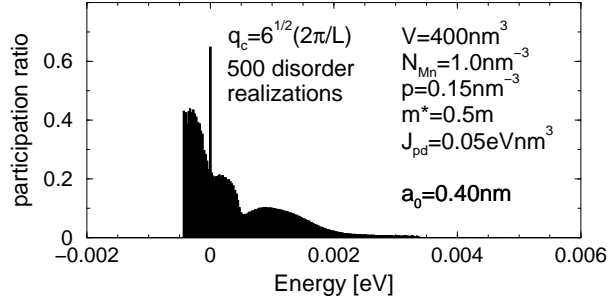}}
\caption{The disorder-averaged participation ratio for the same situation as
in the top panel of Fig.~\protect{\ref{fig_mc_6}}. 
The data is averaged over the
sample intervals of the histogram. The value at zero energy is enhanced 
due to the contribution of the uniform rotation mode in that sample
interval. 
\label{fig_mc_7}}
\end{figure}

To further analyze the nature of this instability we consider the
participation ratios for these excitations which we define by
\begin{equation}
p_j=\left[N_{Mn}V\sum_{I}|\alpha_{I}^{j}(E)|^{4}\right]^{-1}
\label{partrat}
\end{equation}
where $\alpha_{I}^j$ is the $I$-th component of the $j$-th normalized
eigenvector of $D^{-1}_{IJ}(0)$. The ratio $p_j$ is an estimate
for the fraction of Mn sites that have important involvement
in the $j$-th spin wave. For example the zero-mode, uniform
rotation of all spins, has $p_j=1$.
Fig.\ref{fig_mc_7} shows the disorder-averaged participation ratio as
a function of spin-wave energy for the same situation as in 
Fig.\ref{fig_mc_6}. The property that negative energy excitations have large
participation ratios shows that the instabilities of the collinear state
involve correlated reorientations of many spins, rather than lone
loosely coupled moments.

Fig.~\ref{fig_mc_7} shows the 
disorder-averaged participation ratio for the same 
situation as in the top panel of Fig.~\ref{fig_mc_6}.
The negative-energy modes have clearly higher participation ratio than
the eigenvectors at positive energy. This shows that {\em the instability of 
collinear state is due to long-ranged fluctuations involving a large fraction
of the spins present in the system}. Qualitatively the same observations
are made for other values of system parameters.

The effect of a weak external magnetic field on the spin-wave excitation 
spectrum of the noncollinear state is particularly simple. The field 
couples to the local moment through its Land\'e g-factor, 
adding $2 \mu_B H_{ext}$ to the energy of a spin-wave
excitation for $S=5/2$, and to the spin and orbital degrees of freedom 
of the band electrons.  The orbital coupling leads to Landau levels 
that do not play an important role at weak fields in these highly disordered 
samples. Zeeman coupling is 
also unimportant, yielding a contribution to the spin-splitting that is 
negligible compared to the mean-field splitting 
$\Delta=J_{pd}N_{Mn}S\sim 0.1 {\rm eV}$.
It follows that the most negative eigenvalue  of $D^{-1}_{IJ}$ is the 
value of $g_L \mu_B H_{ext}$ necessary to force full spin alignment of a 
noncollinear state. The experiments of Potashnik {\it et al.} 
\cite{Schiffer} demonstrate 
that the maximum value of $M(T=0)$ is achieved over a certain range of 
annealing histories. We associate this maximum value with
the fully aligned collinear Mn configuration state; indeed the maximum 
moment per Mn is consistent with full alignment partially compensated by 
band electrons.  For other annealing histories, $M(T=0)$ is reduced, 
corresponding to noncollinear order of the Mn spins.   
Calculations like those described above show that these spins gradually 
align as an external field is added to the Hamiltonian.  We expect full 
alignment to be indicated experimentally by a kink in the 
$M(T=0,H_{ext})$ curve.  At this point, we predict that
the system will still have gapless excitations and power-law temperature 
dependence of the magnetization, in sharp contrast to the gapful excitations 
and exponentially suppressed temperature dependence that holds for 
conventional ferromagnets in an external
magnetic field. Spin-resonance experiments will nevertheless see a gaped 
spin-wave spectrum in this regime, since they couple only to the zero mode.

Finally we comment on the degree of spin alignment in the noncollinear state.
We generalize our formalism by expanding around and defining
Holstein-Primakoff bosons with respect to a spin coherent state
configuration with orientations 
$\hat\Omega_{I}=(\sin\vartheta_{I}\cos\varphi_{I},
\sin\vartheta_{I}\sin\varphi_{I},\cos\vartheta_{I})$. 
In this general case the  effective action
has a static contribution
(at zero Matsubara frequency only)  
{\em linear} in the Holstein-Primakoff variables:
\begin{equation}
{\cal S}_{\rm fluc}=\frac{1}{2}\sum_{I}
\left[\bar g_{I}z_{I}+g_{I}\bar z_{I}\right]
\label{gradientfluc}
\end{equation}
with $g_{I}=g^{1}_{I}+ig^{2}_{I}$, and
\begin{eqnarray}
g^{1}_{I} & =  & \sqrt{2S}\left(\vec e_{\varphi_{I}}\times\vec e_{z}\right)
\cdot\int d^{3}r\Bigl[J(\vec r-\vec R_{I})\nonumber\\
 & & \Bigl(\left(\langle\vec s(\vec r)\rangle\cdot\vec e_{\varphi_{I}}\right)
\vec e_{\varphi_{I}}+
\left(\langle\vec s(\vec r)\rangle\cdot\vec e_{z}\right)\vec e_{z}
\Bigr)\times\vec \Omega_{I}\Bigr]\\
g^{2}_{I} & =  &\sqrt{2S}\vec e_{z}\cdot
\left(\vec e_{\varphi_{I}}\times\int d^{3}rJ(\vec r-\vec R_{I})
\langle\vec s(\vec r)\rangle\right)
\end{eqnarray} 
where $\vec e_{\varphi_{I}}=(\cos\varphi_{I},\sin\varphi_{I},0)$ and
$\vec e_{z}=(0,0,1)$.

The components $g_{I}$ represent the gradient of the
energy with respect to distortions parametrized by the $z_{I}$. As above in
the case of fluctuations around the collinear state, this zero-frequency
contribution can also be obtained via perturbation theory around the given
state of Mn spin orientations.
The contribution to ${\cal S}_{\rm fluc}$ quadratic in the Holstein-Primakoff
variables involves again a proper Matsubara sum and can be obtained 
straightforwardly. However, its concrete form for the general case 
is more complicated, and shall not be analyzed here.
A given orientation of Mn spins is stationary with respect to fluctuations
if all complex coefficients $g_{I}$ vanish. This is the case if and only if
$\int d^{3}rJ(\vec r-\vec R_{I})\langle\vec s(\vec r)\rangle$ 
is collinear with the direction 
$\hat\Omega_{I}$ of the local ion spin. Therefore, the collinear ferromagnetic
state is always a stationary (but not necessarily stable) spin state.

We have employed the energy gradient expression (\ref{gradientfluc}) in a 
numerical steepest descent procedure to search for true energy minima. 
Our results are as follows: In cases where the energy minimum found by this
method is close to the collinear state (with a magnetization of about
90\% of the maximum value or more), this minimum appears to be
unique for each disorder realization. We can therefore be confident
that we have located 
the absolute ground state of the system. In situations 
where the magnetization is reduced more substantially, however,
(by $\sim 20$\%
 or more) we converge to different
energy minima from different starting points.
In these cases the
model has substantial spin-glass character with a complex energy
landscape. This situation occurs typically for 
larger  ratios
$p/N_{Mn}$. For the system shown in Fig.~\ref{fig_mc_7}, for
instance,  magnetization values at local 
energy minima are typically 
30 to 40\% of the collinear state value.

%% file: concl.tex
\section{Concluding remarks}
\label{section_end}

III-V compound semiconductors with Mn substituted for a small fraction of 
the cations are ferromagnets.  In this chapter we have primarily 
discussed the picture of the physical properties of these materials 
that follows from a model in which each Mn produces a S=5/2 local 
moment and acts as an acceptor.  The valence band holes and the 
exchange interaction that couples them to the local moments are 
both treated phenomenologically.  In the simplest approximation the 
hole bands are taken to be those of the host semiconductor, so that 
this model has a single unknown parameter, the exchange coupling strength,
whose value can be determined experimentally.  Many physical 
properties of this model are relatively simply calculated when a 
virtual crystal approximation is adopted.  When an envelope function
approach is used to describe the semiconductor bands this is equivalent to 
replacing the Mn ions by a continuum, completely eliminating disorder.
We have shown that the mean-field theory of this model predicts critical temperatures
in good agreement with experiment, that it can describe qualitative features of the 
magnetic anisotropy energy including the effect of lattice-matching strains between
the magnetic epilayer and its substrate, and that it implies values of the 
anamolous Hall effect often used to characterize these ferromagnets that are in
good agreement with experiment.  We have also described a theory of the elementary
collective spin-wave excitations of the virtual crystal model.  By comparing
the energy of these elementary excitations with those of the mean-field theory,
we are able to judge the reliability of mean-field-theory critical temperature 
estimates.  In this way we find that we reach the wrong conclusion about mean-field
theory when we make the apparently innocent choice
of using a single parabolic band whose mass is equal to the heavy hole mass
of the host semiconductor.  For such a model, we find that with typical parameters
mean-field theory would give a rather poor estimate of the critical temperature.
However when we use a realistic band model with heavy and light holes that 
are coupled the typical energy of collective magnetic excitations increases
substantially.  The fact that heavy hole states are given approximately by 
J=3/2 spin coherent states with a $\vec k$ dependent orientation, plays an
essential role in producing the enhanced spin stiffness.  {\em We conclude that 
a realistic treatment of the valence band, including its characteristic and 
strong spin-orbit mixing, is absolutely necessary to achieve even a qualitative 
understanding of critical temperature trends in these ferromagnets.} 

Some properties of (III,Mn)V ferromagnets are misrepresented by the 
virtual crystal approximation.   We have examined the effect of Mn site disorder on
these ferromagnets both by performing Monte Carlo calculations and by 
evaluating the collective excitations of disordered systems.  The 
Monte Carlo calculations evaluate the properties of the model exactly
for finite systems, except for treating the Mn spin orientations as classical
which will have a small effect on thermal magnetization suppression at low temperatures.
These calculations confirm conclusions reached by using the virtual crystal approximation
for the most part, but also highlight some of its limitations.  In particular, 
while the collinear magnetic state is always stable in the virtual crystal 
approximation, it can be unstable when disorder is accounted for, leading to non-collinear 
but still ferromagnetic states.  This property provides an 
explanation to the dependence of a given sample's transition temperature
and of the magnetization on 
details of growth and post-growth annealing procedures.

Research on Mn-doped semiconductors is a very active
area of physics, both theoretically and experimentally and there are
many  topics which we have not been able to even touch upon.
For example, studies of {\em nanostructured} semiconductor systems
with Mn-doped components represent an important component of  
developments that are aimed towards future spintronic devices.
Nanostructures such as magnetic quantum wells are expected to be
unusual ferromagnets because of the possibility of using
confinement effects and doping profiles to manipulate their
magnetic properties. The transition temperature
of these systems can be tuned by external electric field \cite{leeprb00},
as demonstrated in recent studies of (In,Mn)As field effect transistors \cite{ohnonature00}.
It is also predicted \cite{lee0106536} that hysteresis properties of magnetic
quantum wells will be extremely sensitive to external bias voltages.
As a result, the magnetization orientation in quantum
wells can be manipulated electrically without changing the
magnetic field.  The possibilities for nano-engineering of material properties
have already been made apparent by the relatively simple (Ga,Mn)As digital ferromagnetic
heterostructures \cite{crookerprl95}.
These systems are grown by incorporating submonolayer planes
of MnAs into a GaAs host. They show ferromagnetic transition
temperatures up to 50 ~K \cite{kawakamiapl00,luoep2ds14}, 
the anomalous Hall effect \cite{luoep2ds14} and remarkably square hysteresis loops
with higher coercivities \cite{kawakamiapl00} than their random alloy
counterparts. A confident modeling of these alternative ferromagnetic
semiconductors with controlled Mn distribution would significantly
contribute to our understanding of magnetic and transport properties
of (III,Mn)V ferromagnets.

In this Chapter we have highlighted one particular approach to modeling these materials.
We anticipate that other approaches will also bring useful insights.  In particular
dynamic mean-field-theory calculations \cite{millis} are able to capture some of the 
Mn alloy disorder physics in a relatively simple way, although they do not 
capture spatial correlations and would not capture non-collinear states.
Similarly first principles calculations \cite{parkphysb00} are likely to prove extremely
valuable in the future, once LDA+U or dynamical-mean-field theory corrections have
been applied to treat the Mn d-electron local moments more accurately.

%% file: acknow.tex
\section*{Acknowledgements}

The work presented in this 
chapter was done in collaboration with a  number of researchers.
M.~Abolfath,
W.A.~Atkinson, J.~Brum, and Byounghak Lee made important contributions
to the mean-field theory of ferromagnetic critical temperature
and magnetic anisotropy in DMS's,  H.H.~Lin has been active in the development
of the spin wave theory, Qian Niu's expertise on the semiclassical transport
theory has been essential for the 
anomalous Hall effect study,
J.~Sinova has been working on correlation effects in 
the itinerant carrier system, and S.-R.E.~Young contributed to the
study of disordered DMS's.
We thank D.D.~Awschalom, D.V.~Baxter, R.N.~Bhatt, A.~Burkov, A.~Chattopadhyay,
A.I.~Chudnovskiy, T.~Dietl, J.~Fernandez-Rossier,
J.~Furdyna,  E.G.~Gwinn, N.A.~Hill, A.J.~Millis, H.~Ohno, F.~von~Oppen, N.~Samarth, 
D.~Sanchez, S.~Sanvito, S.~Das~Sarma, P.~Schiffer, M. van Schilfgaarde, and C.~Timm
for many useful discussions.
The work was supported by the Indiana 21st Century
Fund, the Welch Foundation, DARPA, DOE,
Deutsche Forschungsgemeinschaft, the EU COST program, and the Minsitry
of Education of the Czech Republic.

%% file: book.bbl
\begin{thebibliography}{99.}
\addcontentsline{toc}{section}{References}

\bibitem{ohnosci98} 
H. Ohno, Science {\bf 281}, 951 (1998);
F. Matsukura, H. Ohno, A. Shen, and Y. Sugawara, 
Phys. Rev. B {\bf 57}, R2037 (1998).

\bibitem{kikkawanature99}
J.M. Kikkawa and D.D. Awschalom, Nature {\bf 397}, 139 (1999);
D.D. Awschalom and N. Samarth,
J. Magn. Magn. Mater. {\bf 200}, 130, (1999);
I. Malajovich, J.J. Berry, N. Samarth, and D.D. Awschalom, 
Nature {\bf 411}, 770 (2001).

\bibitem{metals}
J.M. Daughton, J. Appl. Phys. {\bf 81}, 3758 (1997);
J.S. Moodera and G. Mathon,
J. Magn. Magn. Mater. {\bf 200}, 248, (1999);
J. Bass and W.P. Pratt,
J. Magn. Magn. Mater. {\bf 200}, 274, (1999);
K. O'Grady and H. Laidler,
J. Magn. Magn. Mater. {\bf 200}, 616, (1999).

\bibitem{sonoda}
S. Sonoda, S. Shimizu, T. Sasaki, Y. Yamamoto, and H. Hori,
cond-mat/0108159. 

\bibitem{ohnojmmm99} 
H. Ohno, J. Magn. Magn. Mater. {\bf 200}, 110 (1999).

\bibitem{manganitereview}
Y. Tokura and Y. Tomioka,
J. Magn. Magn. Mater. {\bf 200}, 1, (1999);
E. Dagotto, T. Hotta, and A. Moreo, Physics Reports {\bf 344}, 1 (2001).

\bibitem{dietlprb01}
T. Dietl, H. Ohno, and F. Matsukura,
Phys. Rev. B {\bf 63}, 195205 (2001).

\bibitem{handbookarticle}
F. Matsukura and T. Dietl, to be published.

\bibitem{szczytkoprb99}
J. Szczytko, A. Twardowski, K. \'Swiatek, M. Palczewska, M. Tanaka, 
T. Hayashi, and K. Ando,
Phys. Rev. B {\bf 60}, 8304 (1999).

\bibitem{linnarssonprb97}
M. Linnarsson, E. Janz{\'e}n, B. Monemar, M. Kleverman, and A. Thilderkvist, 
Phys. Rev. B {\bf 55}, 6938 (1997).

\bibitem{okabayashiprb98}
J. Okabayashi, A. Kimura, O. Rader, T. Mizokawa, A. Fujimori, T. Hayashi,
and M. Tanaka, Phys. Rev. B {\bf 58}, R4211 (1998).

\bibitem{okabayashiprb99}
J. Okabayashi, A. Kimura, T. Mizokawa, A. Fujimori, T. Hayashi,
and M. Tanaka, Phys. Rev. B {\bf 59}, R2486 (1999). 
 
\bibitem{okabayashiprb01}
J. Okabayashi, A. Kimura, O. Rader, T. Mizokawa, A. Fujimori, T. Hayashi,
and M. Tanaka, Phys. Rev. B {\bf 64}, 125304 (2001).

\bibitem{ohno96}
H. Ohno, F. Matsukura, A. Shen, Y. Sugawara,
A. Oiwa, A. Endo, S. Katsumoto, and Y. Iye, in
{\em Proceedings of the 23rd International Conference on the Physics
of Semiconductors, Berlin 1996}, edited by M. Scheffler and
R. Zimmermann (World Scientific, Singapore, 1996), p. 405.

\bibitem{hayashijcg99}
T. Hayashi, H. Shimada, H. Shimizu, and M. Tanaka,
J. Cryst. Growth {\bf 201/202}, 689 (1999).

\bibitem{bscalcs} 
S. Sanvito and N.A. Hill,
Phys. Rev. B {\bf 62}, 15553 (2000); S. Sanvito, P. Ordej{\'o}n, and N.A. Hill,
Phys. Rev. B {\bf 63}, 165206 (2001); S. Sanvito and N.A. Hill,
cond-mat/0108406;
M. van Schilfgaarde and O.N. Mryasov,
Phys. Rev. B {\bf 63}, 233205 (2001).

\bibitem{akaiprl98}
H. Akai, Phys. Rev. Lett. {\bf 81}, 3002 (1998).

\bibitem{parkphysb00}
J.H. Park, S.K. Kwon, and B.I. Min,
Physica B {\bf 281-283}, 703 (2000).

\bibitem{kossutpss76} 
J. Kossut, Phys. Stat. Solidi {\bf 78}, 537 (1976).

\bibitem{furdynasemi88}
J.K. Furdyna and J. Kossut, {\it Diluted Magnetic Semiconductors}, Vol. 25 of
{\it Semiconductor and Semimetals} (Academic Press, New York, 1988).

\bibitem{larsonprb88}
B.E. Larson, K.C. Hass, and H. Ehrenreich,
Phys. Rev. B {\bf 37}, 4137 (1988).


\bibitem{dietlhandbook94}
T. Dietl, {\it Diluted Magnetic Semiconductors}, Vol. 3B of {\it Handbook of
Semiconductors}, (North-Holland, New York, 1994).

\bibitem{bhattrefs} M. Berciu and R.N. Bhatt, 
Phys. Rev. Lett. {\bf 87}, 107203 (2001).

\bibitem{chudnovskiy} A.I. Chudnovskiy, cond-mat/0108396.

\bibitem{laserbook}
W.W. Chow, S.W. Koch, and M. Sargent III,
{\em Semiconductor Laser Physics}, p.179--192 
(Springer-Verlag, Berlin, 1999).

\bibitem{abolfathprb01} 
M. Abolfath, T. Jungwirth, J. Brum, and A.H. MacDonald, 
Phys. Rev. B {\bf 63}, 054418 (2001).

\bibitem{omiyaphyse}
T. Omiya, F. Matsukura, T. Dietl, Y. Ohno, T. Sakon, M. Motokawa, and H. Ohno, 
Physica E {\bf 7}, 976 (2000).

\bibitem{Bhatta:00}
A.K. Bhattacharjee and C. Benoit $\grave { \rm a}$ la Guillaume, Solid State 
Comm. {\bf 113}, 17 (2000).

\bibitem{fedorych} 
O.M. Fedorych, E.M. Hankiewicz, Z. Wilamowski, and J. Sadowski, 
cond-mat/0106227. 

\bibitem{millis} 
A. Chattopadhyay, S. Das Sarma, and A.J. Millis, cond-mat/0106455.

\bibitem{abrikosov} The role of randomness in a variety of models
where spins are coupled to band electrons has been reviewed by
A.A. Abrikosov, Advances in Physics {\bf 29}, 869 (1980).

\bibitem{jungwirthprb99}
T. Jungwirth, W.A. Atkinson, B.H. Lee, and A.H. MacDonald, 
Phys. Rev. B {\bf 59}, 9818 (1999).

\bibitem{aharoni}
A. Aharoni, {\em Introduction to the Theory of
Ferromagnetism}, (Oxford University Press, New York, 1996).

\bibitem{iiiv}
I. Vurgaftman, J.R. Meyer, and L.R. Ram-Mohan, 
J. Appl. Phys. {\bf 89}, 5815 (2001).

\bibitem{dietlsci00}
T. Dietl, H. Ohno, F. Matsukura, J. Cibert, and D. Ferrand, Science 
{\bf 287}, 1019 (2000).

\bibitem{jungwirthphyse01}
T. Jungwirth and A.H. MacDonald,
Physica E {\bf 10}, 153 (2001).

\bibitem{jungwirthtbp}
T. Jungwirth, J. K\"{o}nig, and A.H. MacDonald,
to be published.

\bibitem{munekataapl93}
H. Munekata, A. Zaslavsky, P. Fumagalli, and R.J. Gambino,
Appl. Phys. Lett. {\bf 63}, 2929 (1993).

\bibitem{skomskicoey} 
R. Skomski and J.M.D. Coey,
{\em Permanent Magnetism} (Institute of Physics Publishing, Bristol, 1999).

\bibitem{jungwniumacdtbp}
T. Jungwirth, Q. Niu, and A.H. MacDonald, cond-mat/0110484.

\bibitem{smitphysica58} 
J. Smit, Physica {\bf 23}, 39 (1958).

\bibitem{bergerprb70}
L. Berger, Phys. Rev. B {\bf 2}, 4559 (1970).

\bibitem{sundaramprb99} 
G. Sundaram and Q. Niu, Phys. Rev. B {\bf 59}, 14915 (1999).

\bibitem{luttingerpr58} 
J.M. Luttinger, Phys. Rev. {\bf 112}, 739 (1958).

\bibitem{ohnoprl92} 
H. Ohno, H. Munekata, T. Penney, S. von Moln\'{a}r, and L.L. Chang, 
Phys. Rev. Lett. {\bf 68}, 2664 (1992).

\bibitem{koenigprl00}
J. K\"onig, H.H. Lin, and A.H. MacDonald, 
Phys. Rev. Lett. {\bf 84}, 5628 (2000); 
Physics E {\bf 10}, 139 (2001).

\bibitem{koenigspringer01}
J. K\"onig, H.H. Lin, and A.H. MacDonald, in {\it Interacting Electrons in 
Nanostructures}, Eds. R. Haug and H. Schoeller, Lecture Notes in Physics 
{\bf 579}, p.195-212 (Springer-Verlag, Berlin, 2001).

\bibitem{schliemannapl01}
J. Schliemann, J. K\"onig, H.H. Lin, and A.H. MacDonald, Appl. Phys. Lett.
{\bf 78}, 1550 (2001).

\bibitem{schliemanncm33}
J. Schliemann, J. K\"onig, and A.H. MacDonald, Phys. Rev. B {\bf 64},
165201 (2001).

\bibitem{koenigcm16}
J. K\"onig, T. Jungwirth, and A.H. MacDonald, 
Phys. Rev. B {\bf 64}, 184423 (2001).

\bibitem{auerbach94}
A. Auerbach, {\em Interacting Electrons and Quantum Magnetism}
(Springer, New York, 1994).

\bibitem{Schiffer}
S.~J. Potashnik, K.~C. Ku, S.~H. Chun, J.~J. Berry, N. Samarth, 
and P. Schiffer, Appl. Phys. Lett. {\bf 79}, 1495 (2001).

\bibitem{Molenkamp}
 G.~M. Schott, W. Faschinger, and L.~W. Molenkamp, cond-mat/0105562. 

\bibitem{schliemanncm0107573}
J. Schliemann and A.H. MacDonald, cond-mat/0107573.

\bibitem{HMC}
S. Duane, A.D. Kennedy, B.J. Pendleton, and D. Roweth, 
Phys. Lett. B {\bf 195}, 216 (1987).

\bibitem{Koenigunpub}
J. K\"onig, J. Schliemann, and A.H. MacDonald, unpublished.

\bibitem{leeprb00}
Byuonghak Lee, T. Jungwirth, and A.H. MacDonald,
Phys. Rev. B {\bf 61}, 15606 (2000).

\bibitem{ohnonature00}
H. Ohno, D. Chiba, F. Matsukura, T. Omiyama, E. Abe, T. Dietl,
Y. Ohno, and K. Ohtani, Nature {\bf 408}, 944 (2000).

\bibitem{lee0106536}
Byounghak Lee, T. Jungwirth, and A.H. MacDonald,
cond-mat/0106536. 

\bibitem{crookerprl95}
S.A. Crooker, D.A. Tulchinsky, J. Levy, D.D. Awschalom,
R. Garcia, and N. Samarth, Phys. Rev. Lett. {\bf 75},
505 (1995).

\bibitem{kawakamiapl00}
R. K. Kawakami, E. Johnston-Halperin, L.F. Chen,
M. Hanson, N. Gu\'ebels, J.S. Speck, A.C. Gossard, and D.D. Awschalom,
Appl. Phys. Lett. {\bf 77}, 2379 (2000).

\bibitem{luoep2ds14}
H. Luo, B.D. McCombe, M.H. Na, K. Mooney, F. Lehmann,
X. Chen, M. Cheon, S.M. Wang, Y. Sasaki, X. Liu, J.K. Furdyna,
Proceedings of the 14the International Conference on the
Electronic Properties of Two-Dimensional Systems, Prague (2001).
 
\end{thebibliography}
